%% file: main.tex
\pdfoutput=1

\documentclass[12pt,a4paper]{article}

\usepackage{ifthen} 
\usepackage{overpic} %
\usepackage[utf8]{inputenc}

\newboolean{pdflatex}
\setboolean{pdflatex}{true} 

\newboolean{articletitles}
\setboolean{articletitles}{true} 

\newboolean{uprightparticles}
\setboolean{uprightparticles}{false} 

\newboolean{highlightchanges}
\setboolean{highlightchanges}{false}

\def\paperauthors{LHCb collaboration} 
\def\paperasciititle{
Test of lepton universality in beauty-quark decays
} %
\def\papertitle{
Test of lepton universality \\ in beauty-quark decays
} %
\def\paperkeywords{{High Energy Physics}, {LHCb}, {Lepton Universality}, {rare decays}} 
\def\papercopyright{\the\year\ CERN for the benefit of the LHCb collaboration} 
\def\paperlicence{CC BY 4.0 licence}
\def\paperlicenceurl{https://creativecommons.org/licenses/by/4.0/}

\input{preamble}
\ifthenelse{\boolean{highlightchanges}}{ \colorlet{changes}{red}}{\colorlet{changes}{black}}

\begin{document}

\renewcommand{\thefootnote}{\fnsymbol{footnote}}
\setcounter{footnote}{1}

\input{title-LHCb-PAPER}


\renewcommand{\thefootnote}{\arabic{footnote}}
\setcounter{footnote}{0}



\pagestyle{plain} 
\setcounter{page}{1}
\pagenumbering{arabic}


%

\input{np_article}

\input{methods}

\input{acknowledgements}

\addcontentsline{toc}{section}{References}
\bibliographystyle{LHCb}
\bibliography{main,standard,LHCb-PAPER,LHCb-CONF,LHCb-DP,LHCb-TDR}

\cleardoublepage
\input{Authorship_LHCb-PAPER-2021-004}

\end{document}

%% file: preamble.tex
\usepackage[top=1in, bottom=1.25in, left=1in, right=1in]{geometry}

\usepackage{microtype}
\usepackage{lineno}  
\usepackage{xspace} 
\usepackage{caption} 

\usepackage{graphicx}  
\usepackage{color}
\usepackage{colortbl}
\graphicspath{{./figs/}} 
\DeclareGraphicsExtensions{.pdf,.PDF,png,.PNG}

\usepackage{amsmath} 
\usepackage{amssymb}
\usepackage{amsfonts}
\usepackage{upgreek} 

\newcommand*\patchAmsMathEnvironmentForLineno[1]{%
\expandafter\let\csname old#1\expandafter\endcsname\csname #1\endcsname
\expandafter\let\csname oldend#1\expandafter\endcsname\csname
end#1\endcsname
 \renewenvironment{#1}%
   {\linenomath\csname old#1\endcsname}%
   {\csname oldend#1\endcsname\endlinenomath}%
}
\newcommand*\patchBothAmsMathEnvironmentsForLineno[1]{%
  \patchAmsMathEnvironmentForLineno{#1}%
  \patchAmsMathEnvironmentForLineno{#1*}%
}
\AtBeginDocument{%
\patchBothAmsMathEnvironmentsForLineno{equation}%
\patchBothAmsMathEnvironmentsForLineno{align}%
\patchBothAmsMathEnvironmentsForLineno{flalign}%
\patchBothAmsMathEnvironmentsForLineno{alignat}%
\patchBothAmsMathEnvironmentsForLineno{gather}%
\patchBothAmsMathEnvironmentsForLineno{multline}%
\patchBothAmsMathEnvironmentsForLineno{eqnarray}%
}

\usepackage{hyperxmp}

\usepackage[pdftex,
            pdfauthor={\paperauthors},
            pdftitle={\paperasciititle},
            pdfkeywords={\paperkeywords},
            pdfcopyright={Copyright (C) \papercopyright},
            pdflicenseurl={\paperlicenceurl}]{hyperref}

\usepackage[colorinlistoftodos,textsize=scriptsize]{todonotes}

\usepackage[all]{hypcap} 

\input{lhcb-symbols-def} 

\usepackage{cite} 
\usepackage{mciteplus}
\usepackage{multirow}
\usepackage{booktabs}

%% file: lhcb-symbols-def.tex
\usepackage{xspace} 
\usepackage{upgreek}

\newcommand{\offsetoverline}[2][0.1em]{\kern #1\overline{\kern -#1 #2}}%

\def\lhcb   {\mbox{LHCb}\xspace}

\def\MagUp {\mbox{\em Mag\kern -0.05em Up}\xspace}

\ifthenelse{\boolean{uprightparticles}}%
{

 \def\Pmu         {\ensuremath{\upmu}\xspace}                 
 \def\Pnu         {\ensuremath{\upnu}\xspace}                 
                  
 \def\Ppi         {\ensuremath{\uppi}\xspace}

 \def\Ptau        {\ensuremath{\uptau}\xspace}

 \def\Ppsi        {\ensuremath{\uppsi}\xspace}

 \def\PDelta      {\ensuremath{\Delta}\xspace}                 
 \def\PXi         {\ensuremath{\Xi}\xspace}                 
 \def\PLambda     {\ensuremath{\Lambda}\xspace}                 
 \def\PSigma      {\ensuremath{\Sigma}\xspace}                 
 \def\POmega      {\ensuremath{\Omega}\xspace}                 
 \def\PUpsilon    {\ensuremath{\Upsilon}\xspace}

 \def\PB      {\ensuremath{\mathrm{B}}\xspace}                 
                  
 \def\PD      {\ensuremath{\mathrm{D}}\xspace}

 \def\PJ      {\ensuremath{\mathrm{J}}\xspace}                 
 \def\PK      {\ensuremath{\mathrm{K}}\xspace}

 \def\Pb      {\ensuremath{\mathrm{b}}\xspace}                 
 \def\Pc      {\ensuremath{\mathrm{c}}\xspace}                 
                  
 \def\Pe      {\ensuremath{\mathrm{e}}\xspace}

 \def\Pi      {\ensuremath{\mathrm{i}}\xspace}

 \def\Pq      {\ensuremath{\mathrm{q}}\xspace}                 
                  
 \def\Ps      {\ensuremath{\mathrm{s}}\xspace}                 
                  
 \def\Pu      {\ensuremath{\mathrm{u}}\xspace}

}
{

 \def\Pmu         {\ensuremath{\mu}\xspace}                 
 \def\Pnu         {\ensuremath{\nu}\xspace}                 
                  
 \def\Ppi         {\ensuremath{\pi}\xspace}

 \def\Ptau        {\ensuremath{\tau}\xspace}

 \def\Ppsi        {\ensuremath{\psi}\xspace}                 
                  
 \mathchardef\PDelta="7101
 \mathchardef\PXi="7104
 \mathchardef\PLambda="7103
 \mathchardef\PSigma="7106
 \mathchardef\POmega="710A
 \mathchardef\PUpsilon="7107
                  
 \def\PB      {\ensuremath{B}\xspace}                 
                  
 \def\PD      {\ensuremath{D}\xspace}

 \def\PJ      {\ensuremath{J}\xspace}                 
 \def\PK      {\ensuremath{K}\xspace}

 \def\Pb      {\ensuremath{b}\xspace}                 
 \def\Pc      {\ensuremath{c}\xspace}                 
                  
 \def\Pe      {\ensuremath{e}\xspace}

 \def\Pi      {\ensuremath{i}\xspace}

 \def\Pq      {\ensuremath{q}\xspace}                 
                  
 \def\Ps      {\ensuremath{s}\xspace}                 
                  
 \def\Pu      {\ensuremath{u}\xspace}

}

\makeatletter
\ifcase \@ptsize \relax
  \newcommand{\miniscule}{\@setfontsize\miniscule{4}{5}}
\or
  \newcommand{\miniscule}{\@setfontsize\miniscule{5}{6}}
\or
  \newcommand{\miniscule}{\@setfontsize\miniscule{5}{6}}
\fi
\makeatother

\DeclareRobustCommand{\optbar}[1]{\shortstack{{\miniscule (\rule[.5ex]{1.25em}{.18mm})}
  \\ [-.7ex] $#1$}}

\def\en         {{\ensuremath{\Pe^-}}\xspace}   
\def\ep         {{\ensuremath{\Pe^+}}\xspace}
\def\epm        {{\ensuremath{\Pe^\pm}}\xspace} 
 
\def\epem       {{\ensuremath{\Pe^+\Pe^-}}\xspace}

\def\mun        {{\ensuremath{\Pmu^-}}\xspace} 

\def\mumu       {{\ensuremath{\Pmu^+\Pmu^-}}\xspace}

\def\taum       {{\ensuremath{\Ptau^-}}\xspace}

\def\ellm       {{\ensuremath{\ell^-}}\xspace}
\def\ellp       {{\ensuremath{\ell^+}}\xspace}
\def\ellell     {\ensuremath{\ell^+ \ell^-}\xspace}

\def\neu        {{\ensuremath{\Pnu}}\xspace}
\def\neub       {{\ensuremath{\overline{\Pnu}}}\xspace}
\def\neue       {{\ensuremath{\neu_e}}\xspace}
\def\neueb      {{\ensuremath{\neub_e}}\xspace}

\def\neulb      {{\ensuremath{\neub_\ell}}\xspace}

\def\quark     {{\ensuremath{\Pq}}\xspace}
\def\quarkbar  {{\ensuremath{\overline \quark}}\xspace}

\def\uquark    {{\ensuremath{\Pu}}\xspace}

\def\squark    {{\ensuremath{\Ps}}\xspace}
\def\squarkbar {{\ensuremath{\overline \squark}}\xspace}

\def\cquark    {{\ensuremath{\Pc}}\xspace}
\def\cquarkbar {{\ensuremath{\overline \cquark}}\xspace}

\def\bquark    {{\ensuremath{\Pb}}\xspace}
\def\bquarkbar {{\ensuremath{\overline \bquark}}\xspace}

\def\pion   {{\ensuremath{\Ppi}}\xspace}

\def\pip    {{\ensuremath{\pion^+}}\xspace}
\def\pim    {{\ensuremath{\pion^-}}\xspace}

\def\kaon    {{\ensuremath{\PK}}\xspace}
  \def\Kbar    {{\kern 0.2em\overline{\kern -0.2em \PK}{}}\xspace}

\def\KorKbar {\kern 0.18em\optbar{\kern -0.18em K}{}\xspace}

\def\Kp      {{\ensuremath{\kaon^+}}\xspace}
\def\Km      {{\ensuremath{\kaon^-}}\xspace}

\def\KS      {{\ensuremath{\kaon^0_{\mathrm{S}}}}\xspace}

\def\Kstarz  {{\ensuremath{\kaon^{*0}}}\xspace}

  \def\Dbar    {{\kern 0.2em\overline{\kern -0.2em \PD}{}}\xspace}
\def\D       {{\ensuremath{\PD}}\xspace}

\def\DorDbar {\kern 0.18em\optbar{\kern -0.18em D}{}\xspace}
\def\Dz      {{\ensuremath{\D^0}}\xspace}
\def\Dzb     {{\ensuremath{\Dbar{}^0}}\xspace}
\def\Dp      {{\ensuremath{\D^+}}\xspace}

\def\Dstarp  {{\ensuremath{\D^{*+}}}\xspace}

\def\Ds      {{\ensuremath{\D^+_\squark}}\xspace}

\def\B       {{\ensuremath{\PB}}\xspace}
\def\Bbar    {{\ensuremath{\kern 0.18em\overline{\kern -0.18em \PB}{}}}\xspace}

\def\BorBbar    {\kern 0.18em\optbar{\kern -0.18em B}{}\xspace}
\def\Bz      {{\ensuremath{\B^0}}\xspace}

\def\Bu      {{\ensuremath{\B^+}}\xspace}

\def\Bp      {{\ensuremath{\Bu}}\xspace}

\def\Bd      {{\ensuremath{\B^0}}\xspace}
\def\Bs      {{\ensuremath{\B^0_\squark}}\xspace}

\def\jpsi     {{\ensuremath{{\PJ\mskip -3mu/\mskip -2mu\Ppsi\mskip 2mu}}}\xspace}
\def\psitwos  {{\ensuremath{\Ppsi{(2S)}}}\xspace}
\def\psiprpr  {{\ensuremath{\Ppsi(3770)}}\xspace}

\def\Y#1S{\ensuremath{\PUpsilon{(#1S)}}\xspace}

\def\Lz          {{\ensuremath{\PLambda}}\xspace}

\def\LorLbar     {\kern 0.18em\optbar{\kern -0.18em \PLambda}{}\xspace}

\def\Lc          {{\ensuremath{\Lz^+_\cquark}}\xspace}

\def\Lb           {{\ensuremath{\Lz^0_\bquark}}\xspace}

\def\BF         {{\ensuremath{\mathcal{B}}}\xspace}
\def\BR         {\BF}

\newcommand{\decay}[2]{\mbox{\ensuremath{#1\!\to #2}}\xspace}         

\def\to                 {\ensuremath{\rightarrow}\xspace}

\def\qsq       {{\ensuremath{q^2}}\xspace}

\def\bsllbar     {\decay{\bquarkbar}{\squarkbar \ell^+ \ell^-}}

\def\btosbar {\decay{\bquarkbar}{\squarkbar}}

\def\btostautaubar {\decay{\bquarkbar}{\squarkbar \tau^+ \tau^-}}

\def\btosmumubar {\decay{\bquarkbar}{\squarkbar \mu^+ \mu^-}}

\def\btoclnubar  {\decay{\bquarkbar}{\cquarkbar \ell^+ {\nu_{\ell}}}}

\def\AT#1     {\ensuremath{A_{\mathrm{T}}^{#1}}\xspace}           

\def\C#1      {\ensuremath{\mathcal{C}_{#1}}\xspace}                       
\def\Cp#1     {\ensuremath{\mathcal{C}_{#1}^{'}}\xspace}                    
\def\Ceff#1   {\ensuremath{\mathcal{C}_{#1}^{\mathrm{(eff)}}}\xspace}        
\def\Cpeff#1  {\ensuremath{\mathcal{C}_{#1}^{'\mathrm{(eff)}}}\xspace}       
\def\Ope#1    {\ensuremath{\mathcal{O}_{#1}}\xspace}                       
\def\Opep#1   {\ensuremath{\mathcal{O}_{#1}^{'}}\xspace}                    

\newcommand{\aunit}[1]{\ensuremath{\text{\,#1}}}       

\newcommand{\tev}{\aunit{Te\kern -0.1em V}\xspace}
\newcommand{\gev}{\aunit{Ge\kern -0.1em V}\xspace}
\newcommand{\mev}{\aunit{Me\kern -0.1em V}\xspace}
\newcommand{\kev}{\aunit{ke\kern -0.1em V}\xspace}
\newcommand{\ev}{\aunit{e\kern -0.1em V}\xspace}
\newcommand{\mevc}{\ensuremath{\aunit{Me\kern -0.1em V\!/}c}\xspace}
\newcommand{\gevc}{\ensuremath{\aunit{Ge\kern -0.1em V\!/}c}\xspace}
\newcommand{\mevcc}{\ensuremath{\aunit{Me\kern -0.1em V\!/}c^2}\xspace}
\newcommand{\gevcc}{\ensuremath{\aunit{Ge\kern -0.1em V\!/}c^2}\xspace}
\newcommand{\gevgevcccc}{\ensuremath{\gev^2\!/c^4}\xspace} 

\def\fb   {\ensuremath{\aunit{fb}}\xspace}
\def\invfb   {\ensuremath{\fb^{-1}}\xspace}

\newcommand{\chisq}{\ensuremath{\chi^2}\xspace}

\newcommand{\chisqip}{\ensuremath{\chi^2_{\text{IP}}}\xspace}

\def\deriv {\ensuremath{\mathrm{d}}}

\def\gsim{{~\raise.15em\hbox{$>$}\kern-.85em
          \lower.35em\hbox{$\sim$}~}\xspace}
\def\lsim{{~\raise.15em\hbox{$<$}\kern-.85em
          \lower.35em\hbox{$\sim$}~}\xspace}

\def\pt         {\ensuremath{p_{\mathrm{T}}}\xspace}

\def\photosplusplus     {\mbox{\textsc{Photos}}\xspace}

\xspace

\def\tell1  {TELL1\xspace}
\def\ukl1   {UKL1\xspace}

\newcommand{\eg}{\mbox{\itshape e.g.}\xspace}
\newcommand{\ie}{\mbox{\itshape i.e.}\xspace}

\def\RK         {\ensuremath{R_{\kaon}}\xspace}
\def\RPsitwos   {\ensuremath{R_{\psitwos}}\xspace}
\def\rjpsi      {\ensuremath{r_{\jpsi}}\xspace}
\def\RH         {\ensuremath{R_{H}}\xspace}
\def\RKstar     {\ensuremath{R_{\Kstarz}}\xspace}

\def\Kll        {\ensuremath{\Kp\ellell}\xspace}

\def\mKee       {\ensuremath{m({\Kp\epem})}\xspace}
\def\mKmm       {\ensuremath{m({\Kp\mumu})}\xspace}
\def\mKll       {\ensuremath{m({\Kp\ellell})}\xspace}

\def\mKllconst  {\ensuremath{m_{\jpsi}{(\Kp\ellell)}}\xspace}

\def\mKllgeneric  {\ensuremath{m_{(\jpsi)}{(\Kp\ellell)}}\xspace}

\def\BuJpsiK  {\decay{\Bu}{\jpsi\Kp}}
\def\Jpsill  {\decay{\jpsi}{\ellell}}

\def\BuJpsiKll  {\decay{\Bu}{\jpsi(\to\ellell)\Kp}}

\def\BuPsiKll  {\decay{\Bu}{\psitwos(\to\ellell)\Kp}}
\def\BuPsiK  {\decay{\Bu}{\psitwos\Kp}}

\def\BuKll  {\decay{\Bu}{\Kp\ellell}}

\def\BuJpsiKmm  {\decay{\Bu}{\jpsi(\to\mumu)\Kp}}
\def\BuPsiKmm  {\decay{\Bu}{\psitwos(\to\mumu)\Kp}}
\def\BuKmm  {\decay{\Bu}{\Kp\mumu}}

\def\BuzHmm  {\decay{\B}{H\mumu}}
\def\BuJpsiKee  {\decay{\Bu}{\jpsi(\to\epem)\Kp}}
\def\BuPsiKee  {\decay{\Bu}{\psitwos(\to\epem)\Kp}}
\def\BuKee  {\decay{\Bu}{\Kp\epem}}

\def\BuzHee  {\decay{\B}{H\epem}}

\def\BuJpsipi  {\decay{\Bu}{\jpsi\pip}}
\def\BdorBu  {{\ensuremath{\B^{(0,+)}}}\xspace}
\def\BdKSll  {\decay{\Bd}{\KS\ellell}}

\def\BuBdKpiplusee {\decay{\BdorBu}{ \Kp\pi^{(-,0)}\epem}}
\def\BuBdKpijpsi {\decay{\BdorBu}{\jpsi(\to\epem)\Kp\pi^{(-,0)}}}

\def\BuDzenu {\decay{\Bu}{\Dzb(\to \Kp\en\neueb)\ep\neue}}

\def\Hc {{\ensuremath{H_c}}\xspace}
\def\HbtoHc {\decay{B}{\Hc(\to\Kp\ellm\neulb X)\ellp\nu_{\ell} Y}}
\def\BuKpipi  {\decay{\Bu}{\Kp\pip\pim}}

\def\RKvalue {\ensuremath{0.846\,^{+\,0.042}_{-\,0.039}\,^{+\,0.013}_{-\,0.012}}}
\def\RKvalueComb {\ensuremath{0.846\,^{+\,0.044}_{-\,0.041}}}

\def\significance {{3.1}\xspace}

%% file: title-LHCb-PAPER.tex

\begin{titlepage}
\pagenumbering{roman}

\vspace*{-1.5cm}
\centerline{\large EUROPEAN ORGANIZATION FOR NUCLEAR RESEARCH (CERN)}
\vspace*{1.5cm}
\noindent
\begin{tabular*}{\linewidth}{lc@{\extracolsep{\fill}}r@{\extracolsep{0pt}}}
\ifthenelse{\boolean{pdflatex}}
{\vspace*{-1.5cm}\mbox{\!\!\!\includegraphics[width=.14\textwidth]{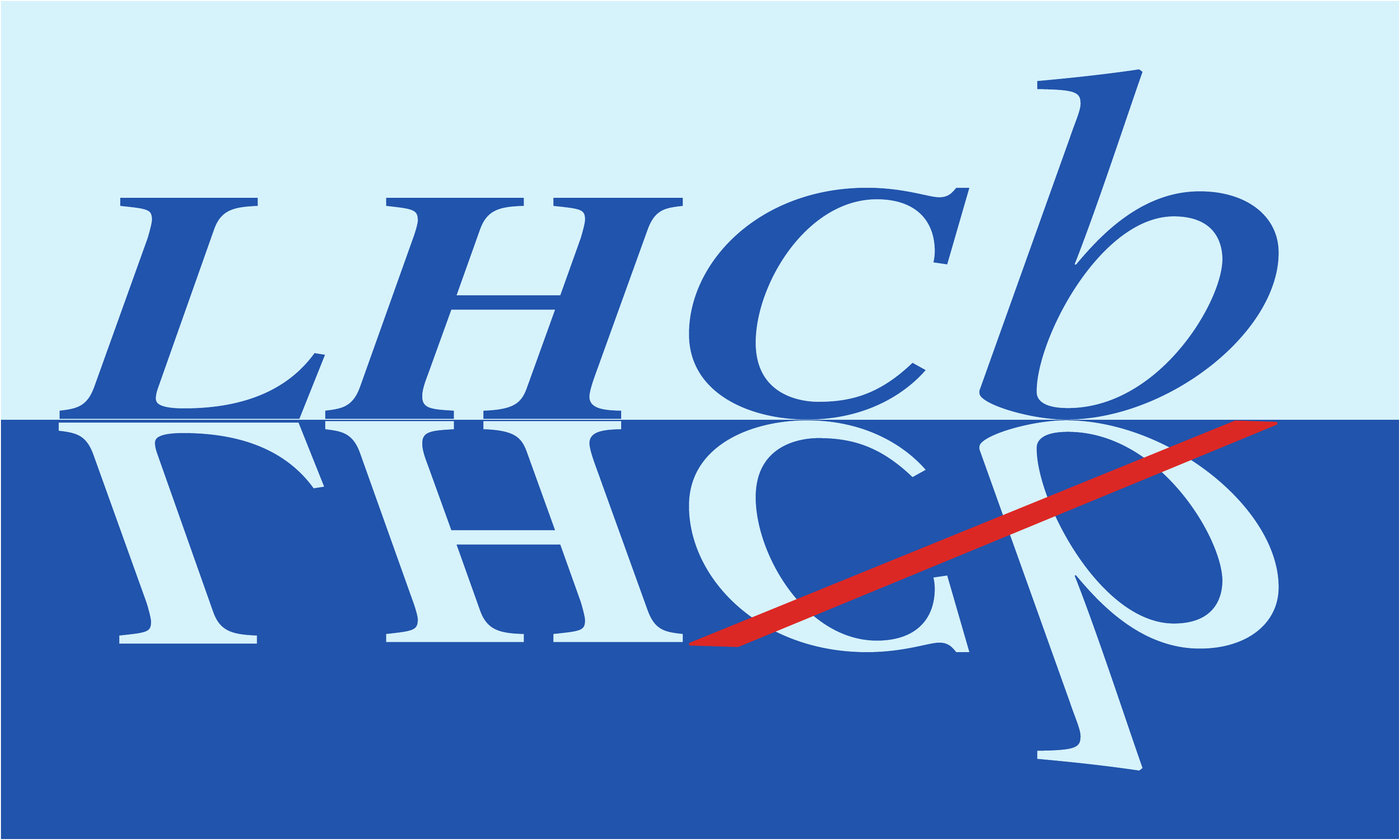}} & &}%
{\vspace*{-1.2cm}\mbox{\!\!\!\includegraphics[width=.12\textwidth]{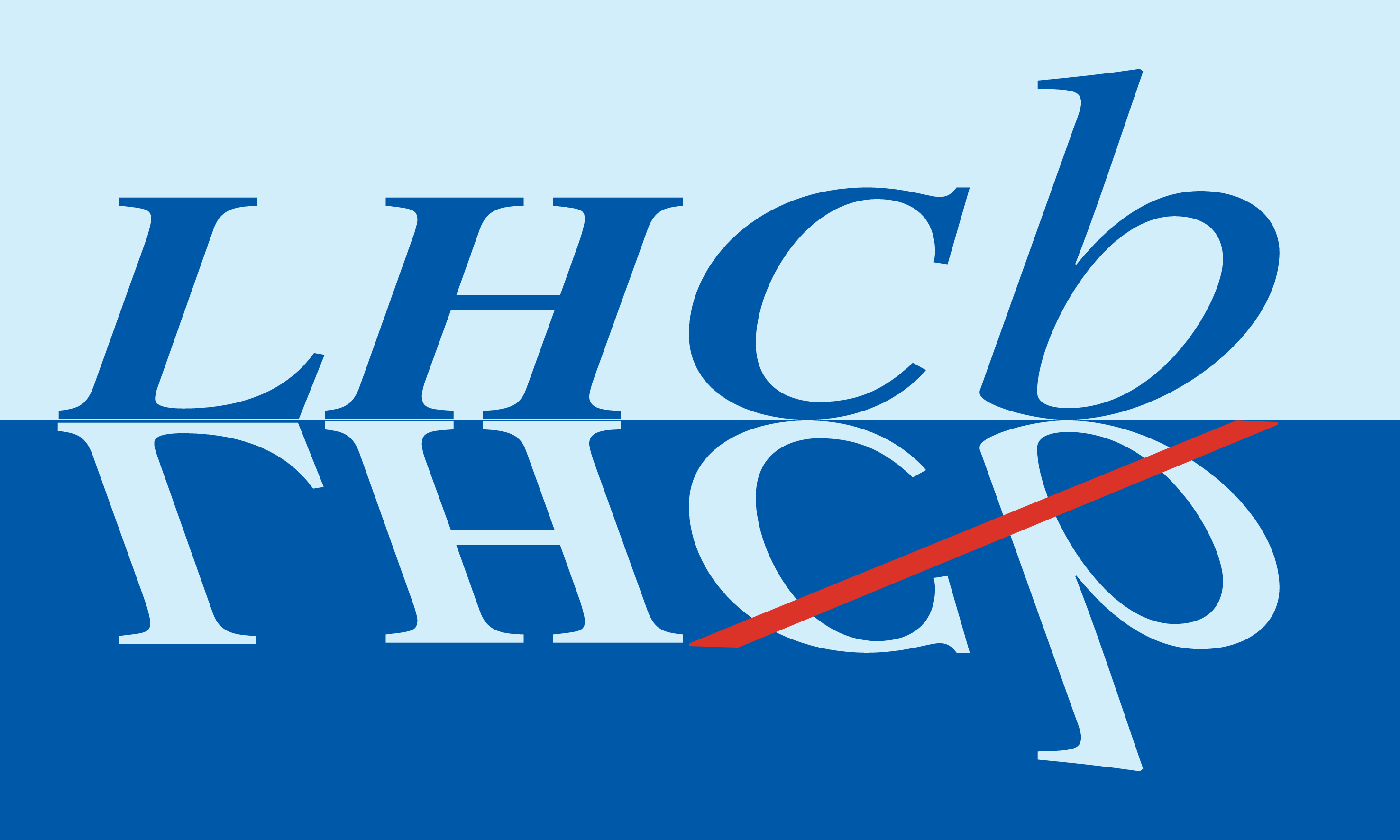}} & &}%
\\
 & & CERN-EP-2021-042 \\  
 & & LHCb-PAPER-2021-004 \\  
 & & 16 March 2022 \\ 
 & & \\
\end{tabular*}

\vspace*{4.0cm}

{\normalfont\bfseries\boldmath\huge
\begin{center}
  \papertitle 
\end{center}
}

\vspace*{2.0cm}

\begin{center}
\paperauthors\footnote{Authors are listed at the end of this paper.}
\end{center}

\vspace{\fill}

\begin{abstract}
  \noindent
The Standard Model of particle physics currently provides our best description of fundamental particles and their interactions. The theory predicts that the different charged leptons, the electron, muon and tau, have identical electroweak interaction strengths. 
Previous measurements have shown a wide range of particle decays are consistent with this principle of lepton universality.
This article presents evidence for the breaking of lepton universality in beauty-quark decays, with a significance of 3.1~standard deviations, based on proton-proton collision data collected with the LHCb detector at CERN’s Large Hadron Collider. 
The measurements are of processes 
in which a beauty meson transforms into a strange meson with the emission of either an electron and a positron, or a muon and an antimuon.
If confirmed by future measurements, this violation of lepton universality would imply physics beyond the Standard Model, such as a new fundamental interaction between quarks and leptons. 

\end{abstract}

\vspace*{2.0cm}

\begin{center}
  Published in Nature Physics 18, (2022) 277-282
\end{center}

\vspace{\fill}

{\footnotesize 
\centerline{\copyright~\papercopyright. \href{\paperlicenceurl}{\paperlicence}.}}
\vspace*{2mm}

\end{titlepage}


\newpage
\setcounter{page}{2}
\mbox{~}

\cleardoublepage

%% file: np_article.tex
The Standard Model (SM) of particle physics provides precise predictions for the properties and interactions of fundamental particles, which have been confirmed by numerous experiments since the inception of the model in the 1960’s. However, it is clear that the model is incomplete. The SM is unable to explain cosmological observations of the dominance of matter over antimatter,  the apparent dark-matter content of the Universe,  or explain the patterns seen in the interaction strengths of the particles.
Particle physicists have therefore been searching for ‘new physics’~\textemdash~the new particles and interactions that can explain the SM’s shortcomings. 

\begin{figure}[b!]
   \begin{center}
      \includegraphics[width=0.96\linewidth]{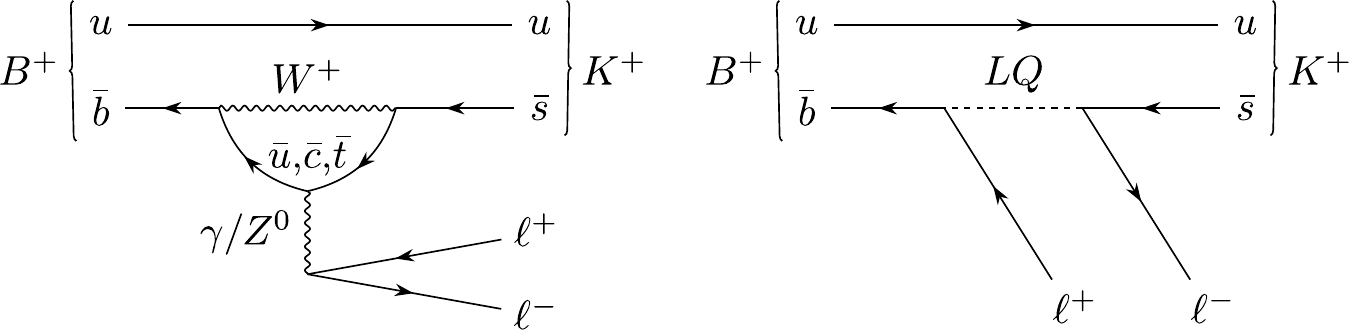}
   \end{center}
   \vspace*{-0.5cm}
   \caption{
  \textcolor{changes}{Contributions to \BuKll decays in the SM and possible new physics models.} A \Bp meson, consisting of \bquarkbar and \uquark quarks, decays into a \Kp, containing \squarkbar and \uquark quarks, and two charged leptons, $\ellp\ellm$. (Left) The SM contribution involves the electroweak bosons $\gamma,~W^+$ and $Z^0$, \textcolor{changes}{and the up-type quarks $\bar{u}$, $\bar{c}$ and $\bar{t}$}. (Right) A possible new physics contribution to the decay with a hypothetical leptoquark ($LQ$) which, unlike the electroweak bosons, could have different interaction strengths with the different types of leptons.     
   }\label{fig:feyn}
\end{figure}

One method to search for new physics is to compare measurements of the properties of hadron decays, where hadrons are bound states of quarks, with their SM predictions. Measurable quantities can be predicted precisely  in the decays 
of a charged beauty hadron, \Bp, into a charged kaon, \Kp, and two charged leptons,~$\ell^{+}\ell^{-}$. The \Bp hadron contains a beauty antiquark, \bquarkbar, and the \Kp a strange antiquark, \squarkbar, such that at the quark level the decay involves a $\bquarkbar\to\squarkbar$ transition.
Quantum field theory allows such a process to be mediated by virtual particles that can have a physical mass larger than the energy available in the interaction. In the SM description of such processes, these virtual particles include the electroweak-force carriers, the $\gamma$, $W^{\pm}$ and $Z^0$ bosons, and the top quark~(see Fig.~\ref{fig:feyn}, left). Such decays are highly suppressed~\cite{Glashow:1970gm} and the fraction of \Bp hadrons that decay into this final state (the branching fraction, \BR) is of the order of $10^{-6}$~\cite{PDG2020}.

A distinctive feature of the SM is that the different leptons, electron ($\en$), muon ($\mun$) and tau ($\taum$),
 have the same interaction strengths. This is known as ‘lepton universality’. The only exception to this is due to the Higgs field, since the lepton-Higgs interaction strength gives rise to the differing lepton masses $m_{\tau}>m_{\mu}>m_e$. 
The suppression of $\bquarkbar \to \squarkbar$ transitions is understood in terms of the fundamental symmetries on which the SM is built. Conversely, lepton universality is an accidental symmetry of the SM, which is not a consequence of any axiom of the theory. Extensions to the SM that aim to address many of its shortfalls predict new virtual particles that could contribute to $\bquarkbar \to \squarkbar$ transitions~(see Fig.~\ref{fig:feyn}, right) and could have nonuniversal interactions, hence giving branching fractions of \BuKll
 decays with different leptons that differ from the SM predictions. Whenever a process is specified in this article, the inclusion of the charge-conjugate mode is implied.

Calculation of the SM predictions for the branching fractions of \BuKmm and \BuKee decays is complicated by the strong nuclear force that binds together the quarks into hadrons, as described by quantum chromodynamics (QCD). The large interaction strengths preclude predictions of QCD effects with the perturbation techniques used to compute the electroweak force amplitudes and only approximate  calculations are presently possible. However, the strong force does not  couple directly to leptons and hence its effect on the \BuKmm and \BuKee decays is identical. The ratio between the branching fractions of these decays is therefore predicted with $\mathcal{O}(1\%)$ precision~\cite{Descotes-Genon:2015uva,Bobeth:2007,Bordone:2016gaq,EOS,Straub:2018kue,Isidori:2020acz}. 
Due to the small masses of both electrons and muons compared to that of \bquark quarks, this ratio is predicted to be close to unity, 
except where the value of the dilepton invariant mass-squared (\qsq) significantly restricts the phase space available to form the two leptons.
Similar considerations apply to decays with other $B$ hadrons, \BuzHmm and \BuzHee, where $B=\Bu$, \Bz, \Bs or \Lb; and $H$ can be \eg an excited kaon, $\Kstarz$, or a combination of particles such as a proton and charged kaon, $p\Km$. 
The ratio of branching fractions, \RH~\cite{Hiller:2003js, Wang:2003je}, is defined in the dilepton mass-squared range $q^{2}_{\rm min} < \qsq < q^{2}_{\rm max}$ as
\begin{equation}
\label{eq:rh}
\RH \equiv  \dfrac{\displaystyle\int_{q^2_\mathrm{min}}^{q^2_{\rm max}} \dfrac{\deriv\BF(\BuzHmm)}{\deriv\qsq} \deriv\qsq}{\displaystyle\int_{q^2_{\rm min}}^{q^2_{\rm max}} \dfrac{\deriv\BF(\BuzHee)}{\deriv\qsq} \deriv\qsq} ~.
\end{equation}
\noindent 
For decays with $H\!=\!\Kp$ and $H\!=\!\Kstarz$ such ratios, denoted \RK and \RKstar, respectively, have previously been measured 
by the LHCb~\cite{LHCb-PAPER-2019-009, LHCb-PAPER-2017-013}, Belle \cite{RKbelle, RKstbelle} and BaBar \cite{RKbabar} collaborations.
For \RK the LHCb measurements are in the region $1.1 < \qsq < 6.0 \gevgevcccc$, whereas for \RKstar the regions are \mbox{$0.045 < \qsq < 1.1 \gevgevcccc$} and $1.1 < \qsq < 6.0 \gevgevcccc$. These ratios have been determined to be 
$2.1$--$2.5$~standard deviations below their respective SM expectations~\cite{Descotes-Genon:2015uva,Bobeth:2007,Bordone:2016gaq,Capdevila:2016ivx,Capdevila:2017ert,Serra:2016ivr,EOS,Straub:2015ica,Straub:2018kue,Altmannshofer:2017fio,Jager:2014rwa,Ghosh:2014awa}. 
The analogous ratio has also been measured for \Lb decays with $H=p\Km$ and is compatible with unity at the level of one standard deviation~\cite{LHCb-PAPER-2019-040}.

These decays all proceed via
 the same \btosbar quark transition and the results have therefore further increased interest in measurements of angular observables~\cite{LHCb-PAPER-2020-041,LHCb-PAPER-2020-002,LHCb-PAPER-2015-051,Aaboud:2018krd,Aubert:2006vb,Lees:2015ymt,Wei:2009zv,Wehle:2016yoi,Aaltonen:2011ja,Khachatryan:2015isa,Sirunyan:2017dhj} and branching fractions~\cite{LHCb-PAPER-2016-012, LHCb-PAPER-2015-023, LHCb-PAPER-2014-006, LHCb-PAPER-2015-009} of decays mediated by \btosmumubar transitions. Such decays also exhibit some tension with the SM predictions but the extent of residual QCD effects is still the subject of debate~\cite{Jager:2014rwa, Descotes-Genon:2015uva,Lyon:2014hpa,Khodjamirian:2012rm,Khodjamirian:2010vf,Descotes-Genon:2014uoa,Horgan:2013pva, Beaujean:2013soa, Hambrock:2013zya, Altmannshofer:2013foa,Bobeth:2017vxj}.
A consistent model-independent interpretation of all these data is possible via a modification of the \mbox{$\bquarkbar \to \squarkbar$} coupling strength~\cite{Ciuchini:2020gvn,Kowalska:2019ley,Alguero:2019ptt,Hurth:2020rzx,Ciuchini:2019usw,Aebischer:2019mlg,Alok:2019ufo}. Such a modification can be realised in new physics models with an additional heavy neutral boson~\cite{Bhattacharya:2014wla,Altmannshofer:2014cfa, Crivellin:2015mga, Celis:2015ara, Falkowski:2015zwa,Belanger:2015nma,Crivellin:2015lwa,Bhattacharya:2016mcc,King:2017anf,Chiang:2017hlj,Falkowski:2018dsl,Allanach:2019mfl,Allanach:2019iiy,Kawamura:2019rth,Dwivedi:2019uqd,Han:2019diw,Capdevila:2020rrl,Altmannshofer:2019xda,Chen:2020szf,Carvunis:2020exc,Karozas:2020zvv,Borah:2020swo,Allanach:2020kss,Sheng:2021tom} or with leptoquarks~\cite{Hiller:2014yaa,Gripaios:2014tna,Varzielas:2015iva,Barbieri:2016las,DiLuzio:2017vat,Crivellin:2017zlb,Becirevic:2017jtw,Greljo:2018tuh,Bordone2018,Fornal:2018dqn,Angelescu:2018tyl,Becirevic:2018afm,Balaji:2019kwe,Cornella:2019hct,Datta:2019tuj,Popov:2019tyc,Bigaran:2019bqv,Bernigaud:2019bfy,DaRold:2019fiw,Fuentes-Martin:2019bue,Hati:2019ufv,Datta:2019bzu,Crivellin:2019dwb,Borschensky:2020hot,Saad:2020ihm,Fuentes-Martin:2020bnh,Dev:2020qet,Fornal:2020ngq,Davighi:2020qqa,Babu:2020hun}. Other explanations of the data involve a variety of extensions to the SM, such as supersymmetry, extended Higgs-boson sectors and models with extra dimensions~\cite{Barman:2018jhz,Blanke:2018sro,Li:2018rax,Shaw:2019fin,Arnan:2019uhr,Trifinopoulos:2019lyo,DelleRose:2019ukt,Ordell:2019zws,Marzo:2019ldg,Darme:2020hpo,Hu:2019ahp,Hu:2020yvs}.
Tension with the SM is also seen in the combination of several ratios that test lepton-universality in \btoclnubar transitions~\cite{Lees:2012xj,LHCb-PAPER-2017-035,Lees:2013uzd,Sato:2016svk,LHCb-PAPER-2015-025,Huschle:2015rga,LHCb-PAPER-2017-027,Belle:2019rba,Hirose:2017dxl}. 
 
In this article, a measurement of the \RK ratio is presented based on proton-proton collision data collected with the LHCb detector at CERN’s Large Hadron Collider (see Methods). The data were recorded during the years 2011, 2012 and 2015--2018, in which the centre-of-mass energy of the collisions was $7$, $8$ and $13\tev$, and correspond to an integrated luminosity of 9\invfb. Compared to the previous LHCb \RK result~\cite{LHCb-PAPER-2019-009}, the experimental method is essentially identical but the analysis uses an additional $4\invfb$ of data collected in 2017 and 2018. The results supersede those of the previous \lhcb analysis. 

The analysis strategy aims to reduce systematic uncertainties induced in modelling the markedly different reconstruction of decays with muons in the final state, compared to decays with electrons. These differences arise due to the significant bremsstrahlung radiation emitted by the electrons and the different detector subsystems that are used to identify electron and muon candidates (see Methods). The major challenge of the measurement is then correcting for the efficiency of the selection requirements used to isolate signal candidates and reduce background. In order to avoid unconscious bias, the analysis procedure was developed and the cross-checks described below performed before the result for \RK was examined. 

In addition to the process discussed above, the \Kll final state is produced via a $\Bp \to X_{\quark\quarkbar}\Kp$ decay, where $X_{\quark\quarkbar}$ is a bound state (meson) such as the \jpsi. The \jpsi meson consists of a charm quark and antiquark, \cquark\cquarkbar, and is produced resonantly at $\qsq=9.59\gevgevcccc$. This ‘charmonium’ resonance subsequently decays into two leptons, \Jpsill. The \BuJpsiKll decays are not suppressed and hence have a branching fraction 
orders of magnitude larger than that of \BuKll decays.
These two processes are separated by applying a requirement on \qsq. The $1.1 < \qsq < 6.0 \gevgevcccc$ region used to select \BuKll decays is chosen to reduce the pollution from the \jpsi resonance and the high-\qsq region that contains contributions from further excited charmonium resonances, such as the \psitwos and \psiprpr states, and from lighter $\squark\squarkbar$ resonances, such as the $\phi(1020)$ meson. In the remainder of this article, the notation \BuKll is used to denote only decays with \mbox{$1.1<\qsq<6.0\gevgevcccc$}, which are referred to as nonresonant, whereas \BuJpsiKll decays are denoted resonant.

To help overcome the challenge of modelling precisely the different electron and muon reconstruction efficiencies, the branching fractions of \BuKll decays are measured relative to those of \BuJpsiK decays~\cite{LHCb-PAPER-2014-024}. Since the $\jpsi\to\ell^+\ell^-$ branching fractions are known to respect lepton universality to within 0.4\%~\cite{Ablikim:2013pqa,PDG2020}, the \RK ratio is determined via the double ratio of branching fractions
    \begin{equation}
    \label{eq:doubleratio}
       \RK = {\frac{\BR(\BuKmm)}{\BR(\BuJpsiKmm)}} \bigg{/} {\frac{\BR(\BuKee)}{\BR(\BuJpsiKee)}} \, .
    \end{equation}
\noindent In this equation, each branching fraction can be replaced by the corresponding event yield divided by the appropriate overall detection efficiency (see Methods), as all other factors needed to determine each branching fraction individually cancel out. 
The efficiency of the nonresonant \BuKee decay therefore needs to be known only relative to that of the resonant \BuJpsiKee decay, rather than relative to the \BuKmm decay. 
As the detector signature of each resonant decay is similar to that of its corresponding nonresonant decay, systematic uncertainties that would otherwise dominate the calculation of these efficiencies are suppressed. The yields observed in these four decay modes and the ratios of efficiencies determined from simulated events then enable an \RK measurement with statistically dominated uncertainties. As detailed below, percent-level control of the efficiencies is verified with a direct comparison of the \BuJpsiKee and \BuJpsiKmm branching fractions in the ratio \mbox{$\rjpsi=\BR(\BuJpsiKmm)/\BR(\BuJpsiKee)$}, which does not benefit from the same cancellation of systematic effects. 

Candidate \BuKll decays are found by combining the reconstructed trajectory~(track) of a particle identified as a charged kaon, together with the tracks from a pair of well-reconstructed oppositely charged particles identified as either electrons or muons. The particles are required to originate from a common vertex, displaced from the proton-proton interaction point, with good vertex-fit quality. The techniques used to identify the different particles and to form \Bp candidates are described in Methods.

The invariant mass of the final state particles, \mKll, is used to discriminate between signal and background contributions, with the signal expected to accumulate around the known mass of the \Bp meson.
Background originates from particles selected from multiple hadron decays, referred to as combinatorial background, and from specific decays of $B$ hadrons. 
The latter also tend to accumulate around specific values of \mKll.
For the muon modes, the residual background is combinatorial and, for the resonant mode, there is an additional contribution from \BuJpsipi decays with a pion misidentified as a kaon. 
For the electron modes, in addition to combinatorial background, other specific background decays contribute significantly in the signal region. The dominant such background for the nonresonant and resonant modes comes from partially reconstructed \BuBdKpiplusee and \BuBdKpijpsi decays, respectively, where the pion is not included in the \Bp candidate. Decays of the form \BuDzenu also contribute at the level of $\mathcal{O}(1\%)$ of the \BuKee signal; and there is also a contribution from \BuJpsiKee decays, where a photon is emitted but not reconstructed. 
The kinematic correlation between \mKee and {\qsq} means that, irrespective of misreconstruction effects, the latter background can only populate the \mKee region well below the signal peak. 

\begin{figure}[!t]
    \centering
    \includegraphics[width=0.45\textwidth]{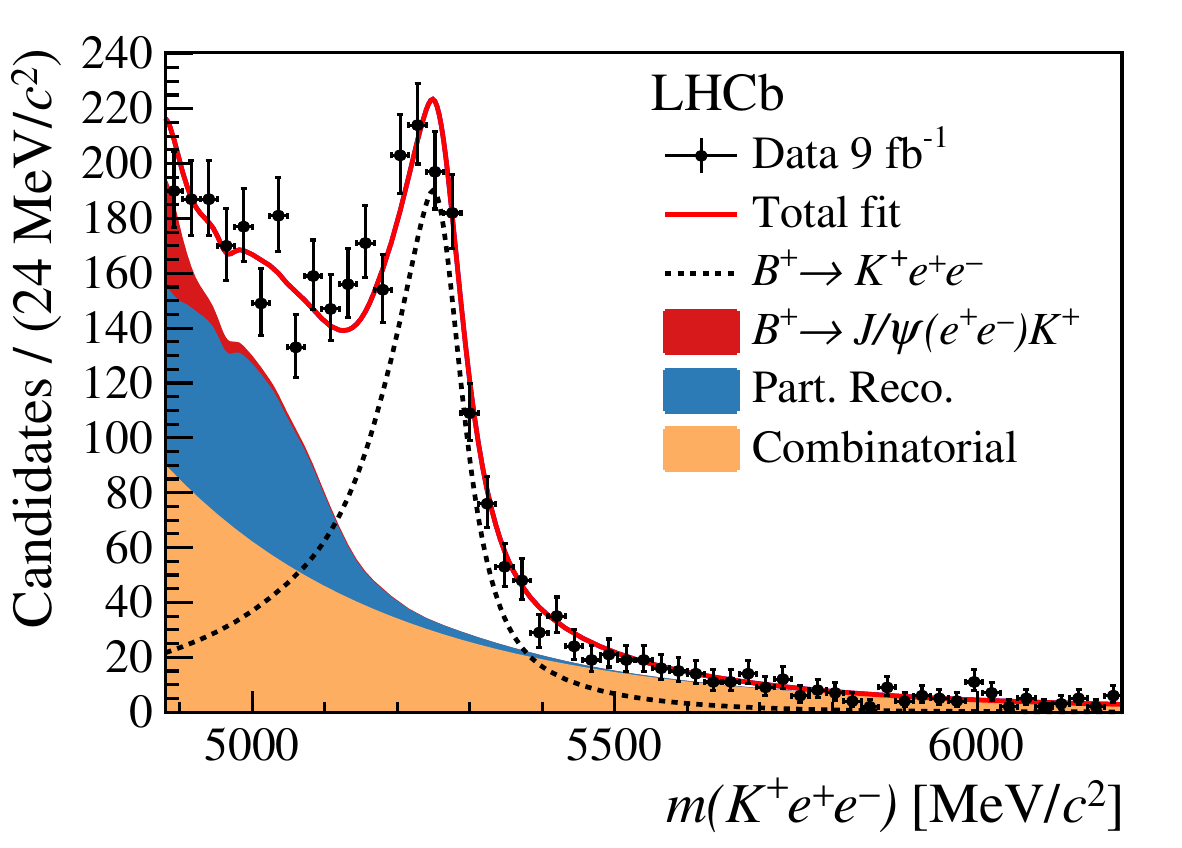}
    \includegraphics[width=0.45\textwidth]{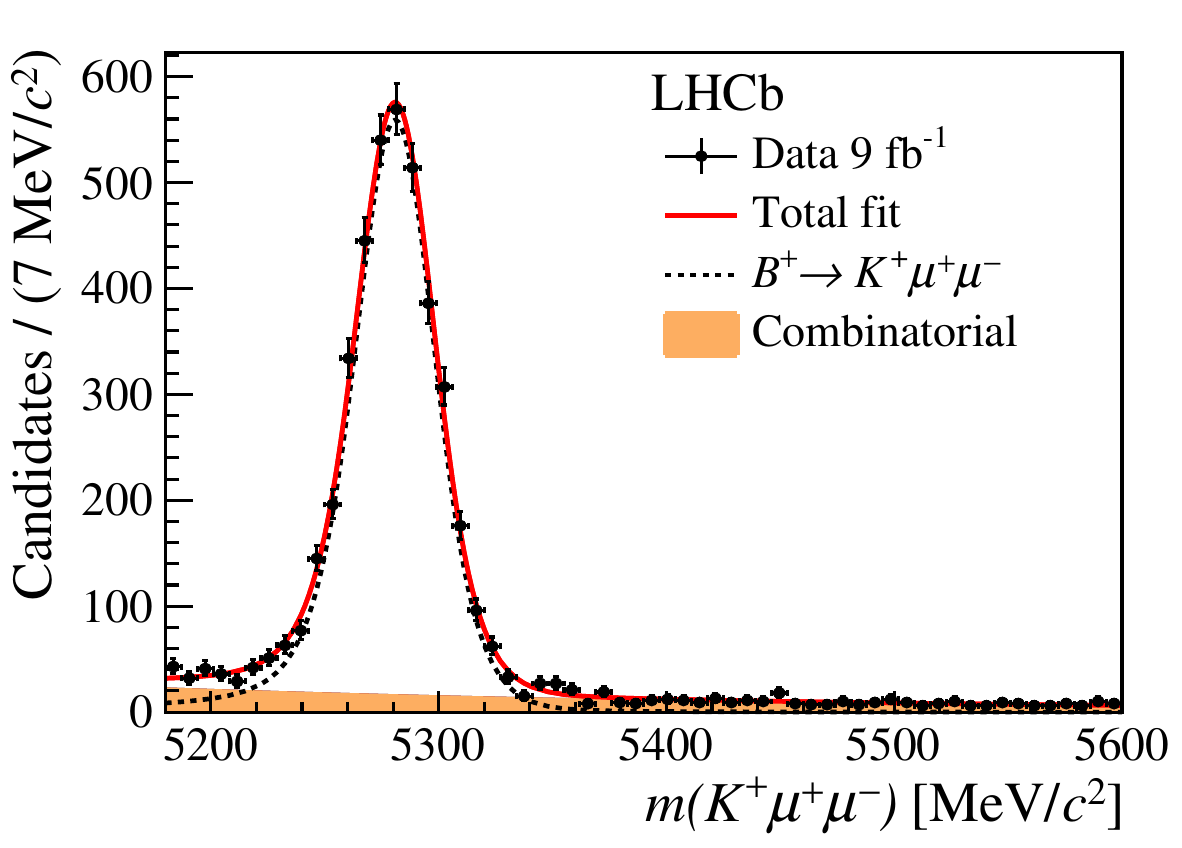}
    \includegraphics[width=0.45\textwidth]{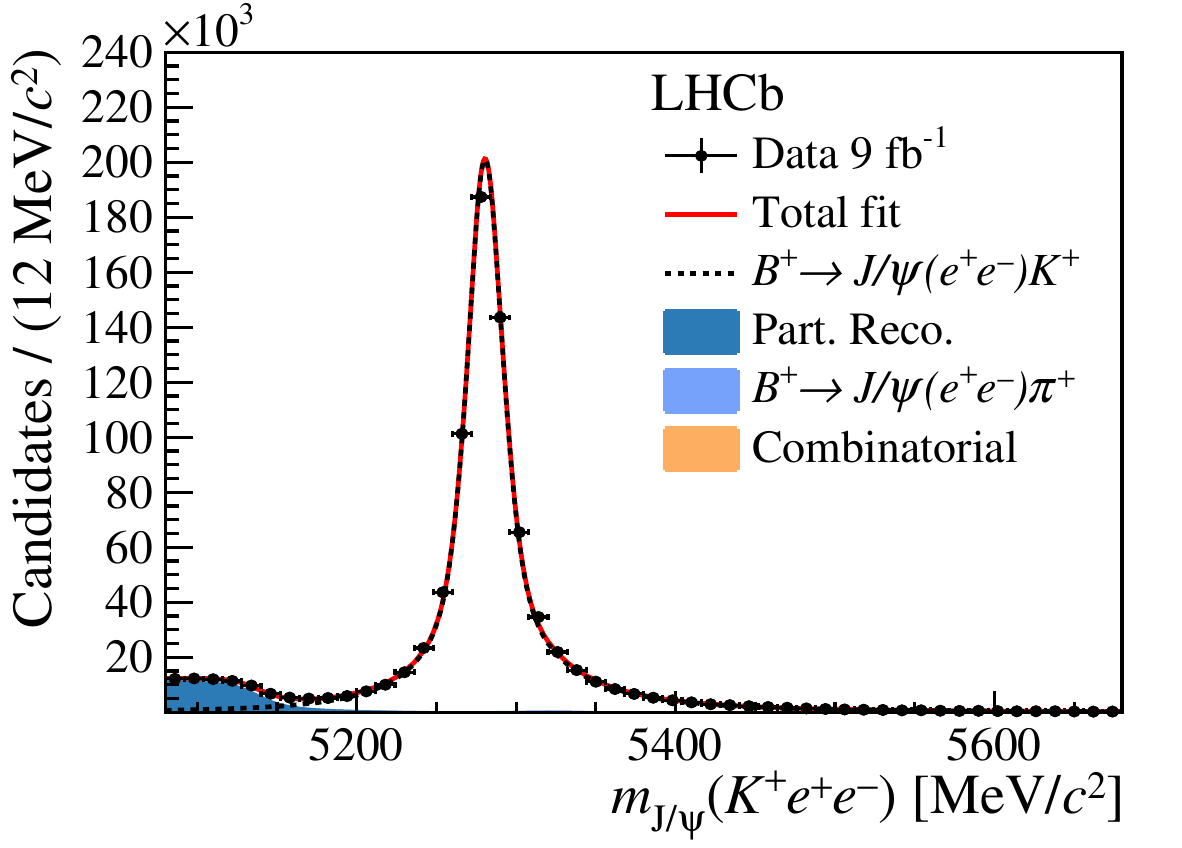}
    \includegraphics[width=0.45\textwidth]{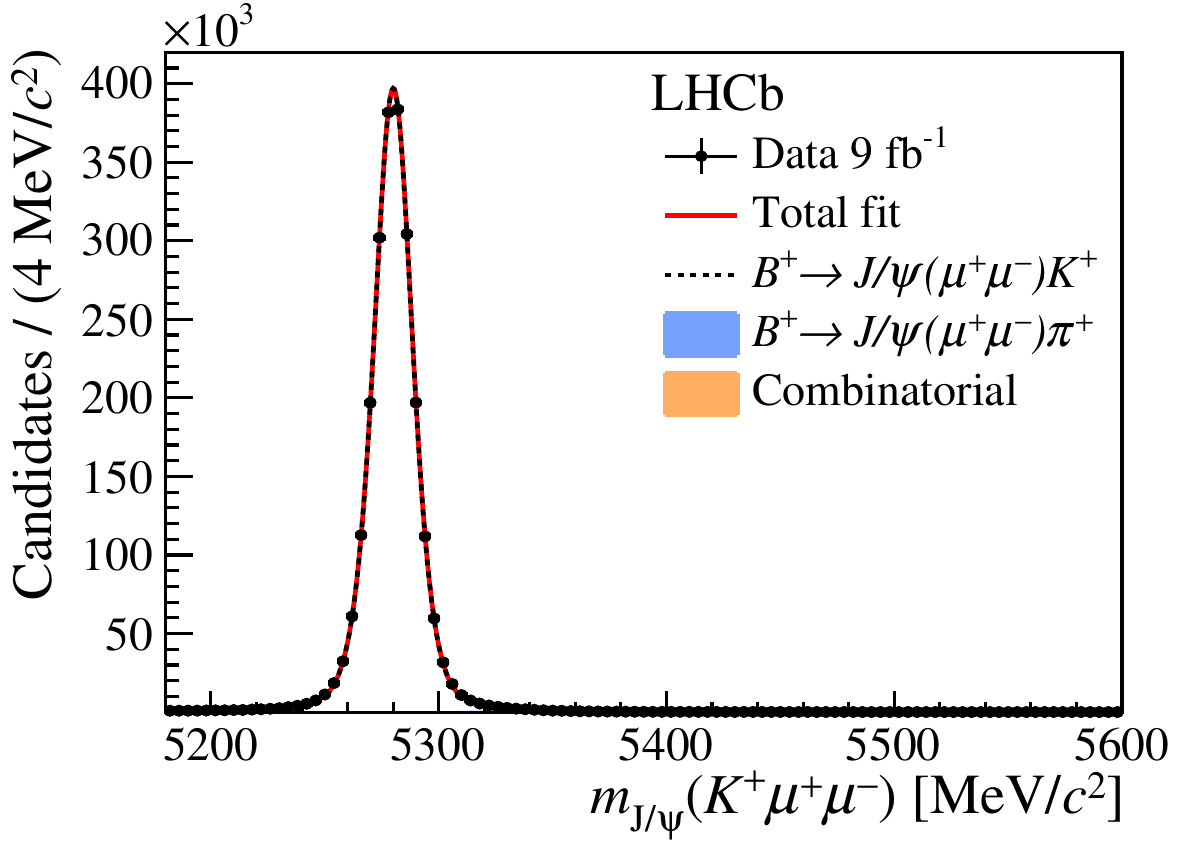}
    \caption{Candidate invariant mass distributions. Distribution of the invariant mass \mKllgeneric for candidates with (left) electron and (right) muon pairs in the final state for the (top) nonresonant \BuKll signal channels and (bottom) resonant \BuJpsiKll decays. The fit projection is superimposed\textcolor{changes}{, with dotted lines describing the signal contribution and solid areas representing each of the background components described in the text and listed in the legend}. In the resonant-mode distributions, some fit components are too small to be visible.}
    \label{fig:fits}
\end{figure}

After the application of the selection requirements, the resonant and nonresonant decays are clearly visible in the mass distributions (see Fig.~\ref{fig:fits}). The yields  in the two \BuKll and two \BuJpsiKll decay modes are determined by performing unbinned extended maximum-likelihood fits to these distributions (see Methods). 
For the nonresonant candidates, the \mKee and \mKmm distributions are fitted with a likelihood function that has the \BuKmm yield and \RK as fit parameters and the resonant decay-mode yields incorporated as Gaussian-constraint terms.
The resonant yields are determined from separate fits
to the mass, \mKllconst, formed by kinematically constraining the dilepton system to the known \jpsi mass~\cite{PDG2020} and thereby improving the mass resolution.

Simulated events are used to derive the two ratios of efficiencies needed to form \RK using Eq.~\eqref{eq:doubleratio}. Control channels are used to calibrate the simulation in order to correct for the imperfect modelling of the \Bu production kinematics and various aspects of the detector response. The overall effect of these corrections on the measured value of \RK is a relative shift of $(+3\pm1)\%$.
When compared with the 20\% shift that these corrections induce in the measurement of \rjpsi, this demonstrates the robustness of the double-ratio method in suppressing systematic biases that affect the resonant and nonresonant decay modes similarly.

The systematic uncertainty (see Methods) from the choice of signal and background mass-shape models in the fits is estimated by fitting pseudoexperiments with alternative models that still describe the data well. The effect on \RK is at the $1\%$ level. A  comparable uncertainty arises from the limited size of the calibration samples, with negligible contributions from the calibration of the \Bu production kinematics and modelling of the selection and particle-identification efficiencies. Systematic uncertainties that affect the ratios of efficiencies influence the measured value of \RK and are taken into account using constraints on the efficiency values. Correlations between different categories of selected events and data-taking periods are taken into account in these constraints. The combined statistical and systematic uncertainty is then determined by scanning the profile-likelihood and the statistical contribution to the uncertainty is isolated by repeating the scan with the efficiencies fixed to their fitted values.  

The determination of the \rjpsi ratio requires control of the relative selection efficiencies for the resonant electron and muon modes, and does not therefore benefit from the cancellation of systematic effects in the double ratio used to measure \RK. Given the scale of the corrections required, comparison of \rjpsi with unity is a stringent cross check of the experimental procedure. 
In addition, if the simulation is correctly calibrated, the measured \rjpsi value will not depend on any variable.
The \rjpsi ratio is therefore also computed as a function of different kinematic variables. 
Even though the nonresonant and resonant samples are mutually exclusive as a function of \qsq, there is significant overlap between them in the quantities on which the efficiency depends, such as the laboratory-frame momenta of the final-state particles, or the opening angle between the two leptons. This is because a given set of values for the final-state particles' momenta and angles in the \Bp rest frame will result in a distribution of such values when transformed to the laboratory frame. 

\begin{figure}[!b]
   \begin{center}

\includegraphics[width=0.45\linewidth,trim={0 0 0 0.5cm}, clip]{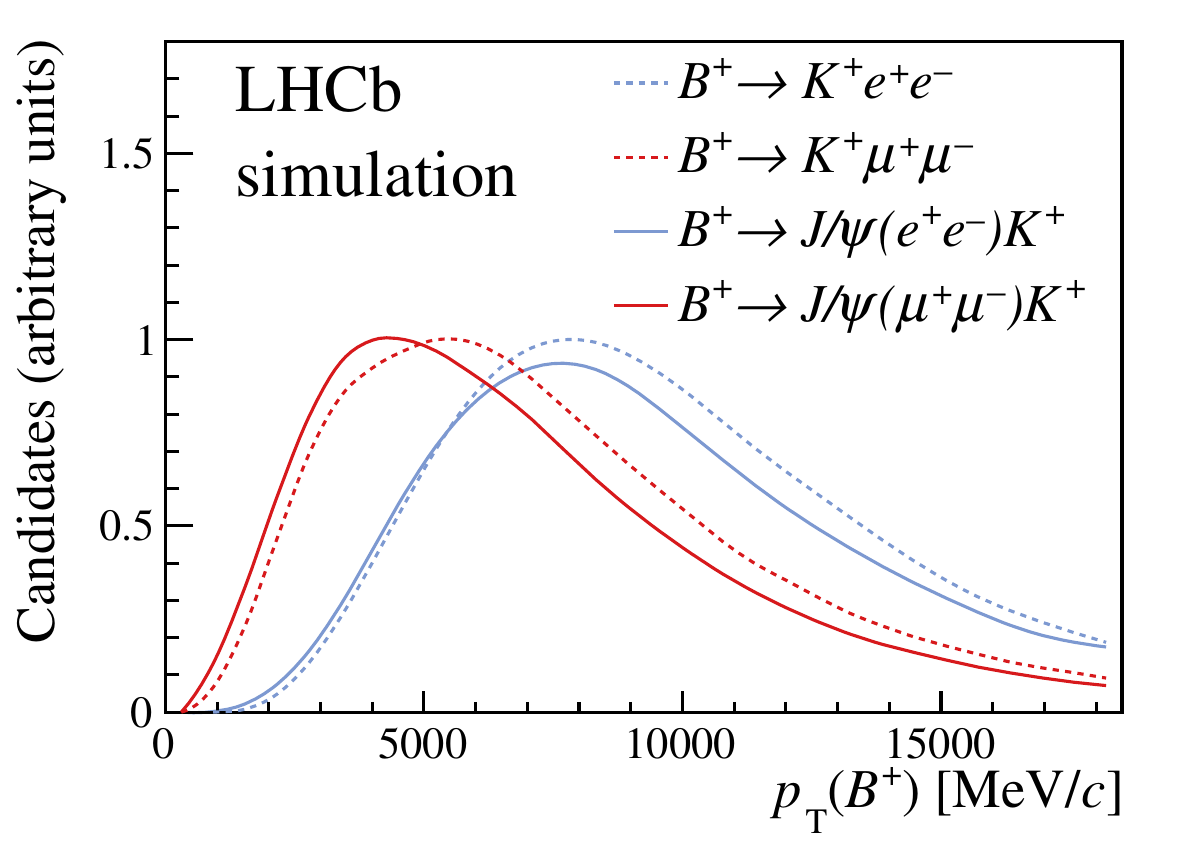}
\includegraphics[width=0.45\linewidth,trim={0 0.15cm 0 0}, clip]{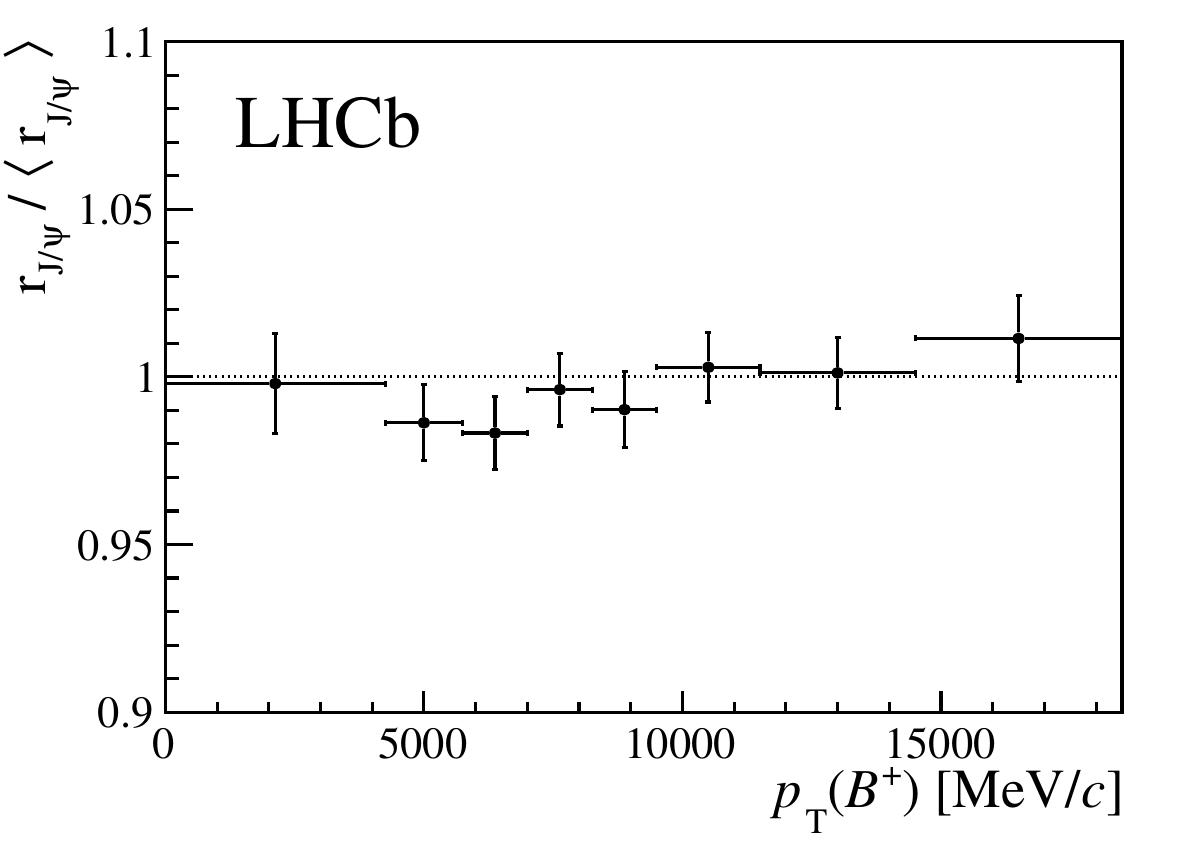}
      \end{center}
      \caption{Differential \rjpsi measurement. The 
      distributions of (left) the \Bp transverse momentum, \pt, and (right) the ratio \rjpsi relative to its average value $\left< \rjpsi \right>$ as a function of \pt. The \pt spectrum of the \BuJpsiK decays is similar to that of the corresponding \BuKll decays such that the measurement of \rjpsi tests the kinematic region relevant for the \RK measurement. The lack of any dependence of the value of $\rjpsi/\left< \rjpsi \right>$ as a function of \Bp $p_{\mathrm T}$ demonstrates control of the efficiencies. \textcolor{changes}{Uncertainties on the data points are statistical only.}}
    \label{fig:rjpsi_differential}
\end{figure}

The value of \rjpsi is measured to be $0.981\pm0.020$. This uncertainty includes both statistical and systematic effects, where the latter dominate. 
The consistency of this ratio with unity demonstrates control of the efficiencies well in excess of that needed for the determination of \RK. 
In the measurement of the \rjpsi ratio, the systematic uncertainty is dominated by the imperfect modelling of the \Bu production kinematics and the modelling of selection requirements, which have a negligible impact on the \RK measurement.
No significant trend is observed in the differential determination of \rjpsi as a function of any considered variable. An example distribution, with \rjpsi determined as a function of \Bp momentum component transverse to the beam direction, \pt, is shown in Fig.~\ref{fig:rjpsi_differential}. Assuming the observed \rjpsi variation in such distributions reflects genuine mismodelling of the efficiencies, rather than statistical fluctuations, and taking into account the spectrum of the relevant variables in the nonresonant decay modes, a total shift on \RK is computed for each of the variables examined. The resulting variations are typically at the permille level and hence well within the estimated systematic uncertainty on \RK. Similarly, 
computations of the \rjpsi ratio in bins of two kinematic variables also do not show any trend and are consistent with the systematic uncertainties assigned on the \RK measurement.

In addition to \BuJpsiK decays, 
clear signals are observed from \BuPsiK decays. 
The double ratio of branching fractions, \RPsitwos, defined by
\begin{equation}
\label{eq:RPsitwos}
\RPsitwos  =
{\frac{\BR(\BuPsiKmm)}{\BR(\BuJpsiKmm)}} \bigg{/} {\frac{\BR(\BuPsiKee)}{\BR(\BuJpsiKee)}}  \,,
\end{equation}
\noindent provides an independent validation of the double-ratio analysis procedure and further tests the control of the efficiencies. 
This double ratio is expected to be close to unity~\cite{PDG2020}
and is determined to be $0.997\pm 0.011$, where the uncertainty includes both statistical and systematic effects, the former of which dominates.
This can be interpreted as a world-leading test of lepton flavour universality in $\psi(2S) \rightarrow \ell^+\ell^-$ decays.

\begin{figure}[!t]
    \centering
    \includegraphics[width=0.7\textwidth]{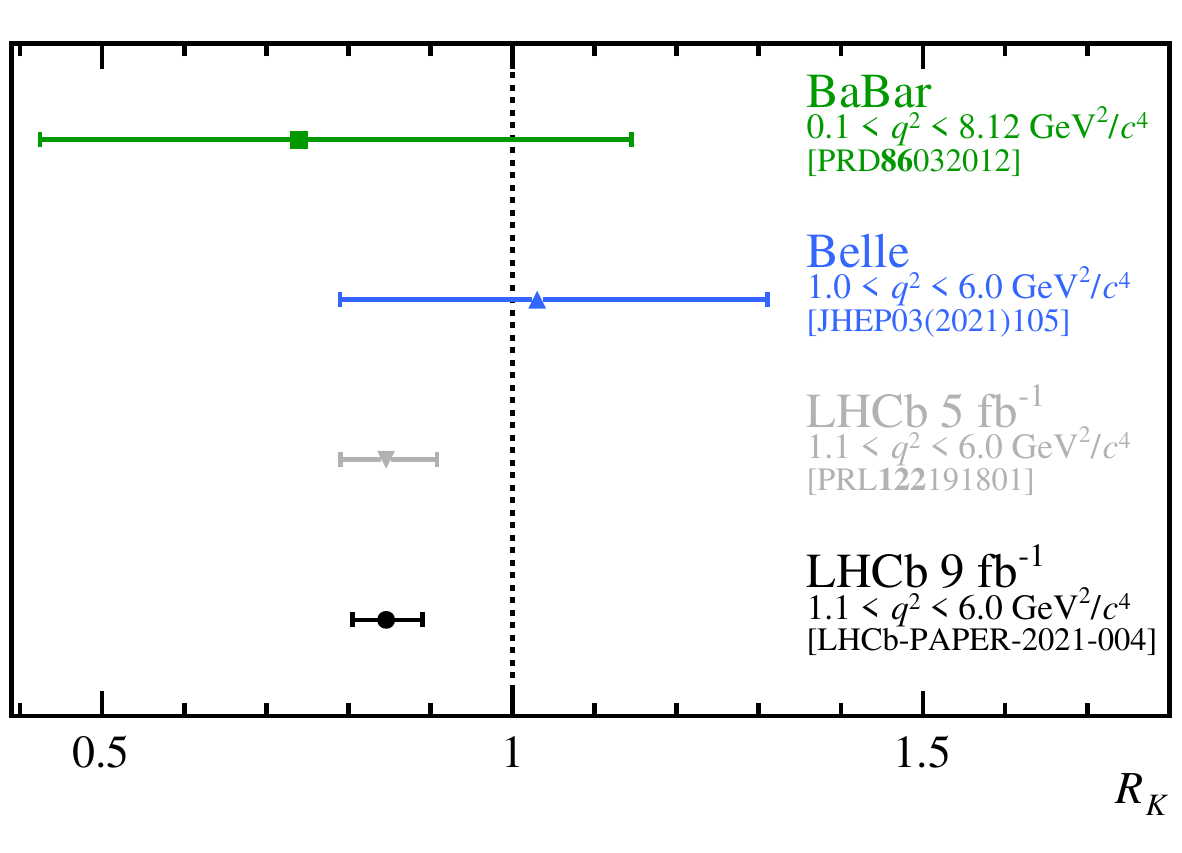} 
    \caption{Comparison between \RK measurements. 
    In addition to the LHCb result, the measurements by the BaBar~\cite{RKbabar} and Belle~\cite{RKbelle} collaborations, which combine \BuKll and \BdKSll decays, are also shown. \textcolor{changes}{The vertical dashed line indicates the SM prediction.}
    }
    \label{fig:RKresult}
\end{figure}

The fit projections for the \mKll and \mKllconst distributions are shown in Fig.~\ref{fig:fits}. The fit is of good quality and the value of \RK is measured to be
\begin{displaymath}
\RK (1.1 < \qsq < 6.0 \gevgevcccc) = \RKvalue\,,
\end{displaymath}
\noindent where the first uncertainty is statistical and the second systematic. Combining the uncertainties gives $\RK=\RKvalueComb$.
This is the most precise measurement to date and is consistent with the SM expectation, \mbox{$1.00 \pm 0.01$}~\cite{Descotes-Genon:2015uva,Bobeth:2007,Bordone:2016gaq,EOS,Straub:2018kue}, at the level of 0.10\% (\significance~standard deviations), giving evidence for the violation of lepton universality in these decays.
The value of \RK is found to be consistent in subsets of the data divided on the basis of data-taking period, different selection categories and magnet polarity (see Methods).
The profile-likelihood is given in Methods. A comparison with previous measurements is shown in Fig.~\ref{fig:RKresult}.

The $3850\pm 70$ \BuKmm decay candidates that are observed are used to compute the \BuKmm  branching fraction as a function of \qsq. The results are consistent between the different data-taking periods and with previous \lhcb measurements~\cite{LHCb-PAPER-2014-006}.
The \BuKee branching fraction is determined by combining the value of \RK with the value of $\deriv\BR(\BuKmm)/\deriv\qsq$ in the region $(1.1 < \qsq < 6.0 \gevgevcccc)$~\cite{LHCb-PAPER-2014-006}, taking into account correlated systematic uncertainties. This gives
\begin{displaymath}
\frac{\deriv\BR(\BuKee)}{\deriv\qsq}(1.1 < \qsq < 6.0 \gevgevcccc)  = (28.6\,^{+\, 1.5}_{-\, 1.4} \,\pm 1.3)\times 10^{-9}\,c^4\kern -0.1em/\kern -0.15em\gev^2\,.
\end{displaymath}
\noindent The 1.9\% uncertainty on the \BuJpsiK branching fraction~\cite{PDG2020} gives rise to the dominant systematic uncertainty. This is the most precise measurement of this quantity to date and, given the large ($\mathcal{O}(10\%)$) theoretical uncertainty on the predictions~\cite{Khodjamirian:2017fxg,Straub:2018kue}, is consistent with the SM.

A breaking of lepton universality would require an extension of the gauge structure of the SM that gives rise to the known fundamental forces. It would therefore constitute a significant evolution in our understanding and would challenge an inference based on a wealth of experimental data in other processes. Confirmation of any beyond the SM effect will clearly require independent evidence from a wide range of sources.

Measurements of other $R_H$ observables with the full LHCb data set will provide further information on the quark-level processes measured. 
In addition to affecting the decay rates, new physics can also alter how the decay products are distributed in phase space. 
An angular analysis of the electron mode, where SM-like behaviour might be expected in the light of the present results and those from \btosmumubar decays, would allow the formation of ratios between observable quantities other than branching fractions, enabling further precise tests of lepton universality~\cite{Altmannshofer:2015mqa,Capdevila:2016ivx,Wehle:2016yoi,Geng:2017svp,Serra:2016ivr}. The hierarchical effect needed to explain the existing \bsllbar and \btoclnubar data, with the largest effects observed in tau modes, then muon modes, and little or no effects in electron modes, suggests that studies of \btostautaubar transitions are also of great interest~\cite{LHCb-PAPER-2017-003, TheBaBar:2016xwe}. There are excellent prospects for all of the above and further measurements with the much larger samples that will be collected with the upgraded LHCb detector from 2022 and, in the longer term, with the LHCb Upgrade~II~\cite{LHCb-PII-Physics}. Other experiments should also be able to determine $R_H$ ratios, with the Belle II experiment in particular expected to have competitive sensitivity~\cite{Kou:2018nap}. The ATLAS and CMS experiments may also be able to contribute~\cite{CMSBparking,ATLASL1topotrig}.

In summary, in the dilepton mass-squared region $1.1 < \qsq < 6.0 \gevgevcccc$, the ratio of branching fractions for \BuKmm and \BuKee decays is measured to be $\RK=\RKvalueComb$. 
This is the most precise measurement of this ratio to date and is compatible with the SM prediction with a p-value of 0.10\%. The significance of this discrepancy is \significance standard deviations, giving evidence for the violation of lepton universality in these decays.

%% file: methods.tex
\newpage
\section*{Methods}
\subsection*{Experimental setup}

The Large Hadron Collider (LHC) is the world’s highest-energy particle accelerator and is situated approximately 100\,m underground, close to Geneva, Switzerland. The collider accelerates two counter-rotating beams of protons, guided by superconducting magnets located around a 27\,km circular tunnel, and brings them into collision at four interaction points that house large detectors. The LHCb experiment~\cite{Alves:2008zz,LHCb-DP-2014-002} is instrumented in the region covering the polar angles between 10 and 250\,mrad around the proton beam axis, in which the products from $B$-hadron decays can be efficiently captured and identified. The detector includes a high-precision tracking system with a dipole magnet, providing measurements of momentum and impact parameter (IP), defined for charged particles as the minimum distance of a track to a primary proton-proton interaction vertex (PV). Different types of charged particles are distinguished using information from two ring-imaging Cherenkov (RICH) detectors, a calorimeter and a muon system~\cite{LHCb-DP-2014-002, LHCb-DP-2014-001, LHCb-DP-2013-003,LHCb-DP-2013-001,LHCb-DP-2012-003,LHCb-DP-2012-002}. 

Since the associated data storage and analysis costs would be prohibitive, the experiment does not record all collisions. 
Only potentially interesting events, selected using real-time event filters referred to as triggers, are recorded.
The LHCb trigger system has a hardware stage, based on information from the calorimeter and muon systems; followed by a software stage that uses all the information from the detector, including the tracking, to make the final selection of events to be recorded for subsequent analysis. The trigger selection algorithms are based on identifying key characteristics of $B$ hadrons and their decay products, such as high \pt final state particles, and a decay vertex that is significantly displaced from any of the PVs in the event.

For the \RK measurement, candidate events are  required to have passed a hardware trigger algorithm that selects either a high \pt muon; or an electron, hadron or photon with high transverse energy deposited in the calorimeters. The \BuKmm and \BuJpsiKmm candidates must be triggered by one of the muons, whereas \BuKee and \BuJpsiKee candidates must be triggered in one of three ways: by either one of the electrons; by the kaon from the \Bp decay; or by particles in the event that are not decay products of the \Bp candidate. In the software trigger, the tracks of the final-state particles are required to form a displaced vertex with good fit quality. A multivariate algorithm is used for the identification of displaced vertices consistent with the decay of a $B$ hadron~\cite{BBDT, LHCb-PROC-2015-018}.

\subsection*{Analysis description}

The analysis technique used to obtain the results presented in this article is essentially identical to that used to obtain the previous LHCb \RK measurement, described in Ref.~\cite{LHCb-PAPER-2019-009} and only the main analysis steps are reviewed here. 

\subsubsection*{Event selection} 

Kaon and muon candidates are identified using the output of multivariate classifiers that exploit information from the tracking system, the RICH detectors, the calorimeters and the muon chambers.
Electrons are identified by matching tracks to  particle showers in the electromagnetic calorimeter~(ECAL) and using the ratio of the energy detected in the ECAL to the momentum measured by the tracking system. 
An electron that emits a bremsstrahlung photon due to interactions with the material of the detector downstream of the dipole magnet results in the photon and electron depositing their energy in the same ECAL cells, and therefore in a correct measurement of the original energy of the electron in the ECAL. However, a bremsstrahlung photon emitted upstream of the magnet  will deposit energy in a different part of the ECAL than the electron, which is deflected in the magnetic field. For each electron track, a search is therefore made in the ECAL for energy deposits around the extrapolated track direction before the magnet that are not associated with any other charged tracks. The energy of any such deposit is added to the electron energy that is derived from the measurements made in the tracker. Bremsstrahlung photons can be added to none, either, or both of the final-state \ep and \en candidates. 

In order to suppress background, each final-state particle is required to have sizeable \pt and to be inconsistent with coming from a PV. The particles are required to originate from a common vertex, with good vertex-fit quality, that is displaced significantly from all of the PVs in the event. The PVs are reconstructed by searching for space points where an accumulation of track trajectories is observed. A weighted least-squares method is then employed to find the precise vertex position. The \Bp momentum vector is required to be aligned with the vector connecting one of the PVs in the event (below referred to as the associated PV) and the \Bp decay vertex. The value of \qsq is calculated using only the lepton momenta, without imposing any constraint on the \mKll mass. 

The \mKll mass ranges and the \qsq regions used to select the different decay modes are shown in Table~\ref{tab:q2ranges}. The selection requirements applied to the nonresonant and resonant decays are otherwise identical. For the muon modes, the superior mass resolution allows a fit in a reduced \mKll mass range compared to the electron modes. For the electron modes, a wider mass region is needed to perform an accurate fit, but the range chosen suppresses any significant contribution from decays with two or more additional pions that are not reconstructed. The residual contribution from such decays is considered as a source of systematic uncertainty. Resolution effects similarly motivate the choice of nonresonant \qsq regions, with a lower limit that excludes contributions from $\phi$-meson decays and an upper limit that reduces the tail from \BuJpsiKee decays. \textcolor{changes}{The proportion of signal candidates that migrate in and out of the \qsq region of interest is of the order of 10\%. This effect is accounted for using simulation.}

\begin{table}[t]
\centering
\caption{Nonresonant and resonant mode $\qsq$ and \mKll ranges. The variables \mKll and \mKllconst are used for nonresonant and resonant decays, respectively.}\label{tab:q2ranges}
\begin{tabular}{ccc}
\toprule
Decay mode & \qsq & \mKllgeneric \\
           & $[\gevgevcccc]$ & $[\gevcc]$ \\
\midrule
{\begin{tabular}{r@{\;}l}nonresonant&$e^+e^-$\\resonant&$e^+e^-$\\nonresonant&$\mu^+\mu^-$\\resonant&$\mu^+\mu^-$\end{tabular}}     & {\begin{tabular}{r@{.}l@{\;--\;}r@{.}l}1&1 & 6&0\\6&00 & 12&96\\1&1 & 6&0\\8&68&10&09\end{tabular}}    & {\begin{tabular}{r@{\;--\;}l}4.88&6.20\\5.08&5.70\\5.18&5.60\\5.18&5.60\end{tabular}}\\
\bottomrule
\end{tabular}
\end{table}

\begin{figure}[!tb]
   \begin{center}
      \includegraphics[width=0.48\linewidth]{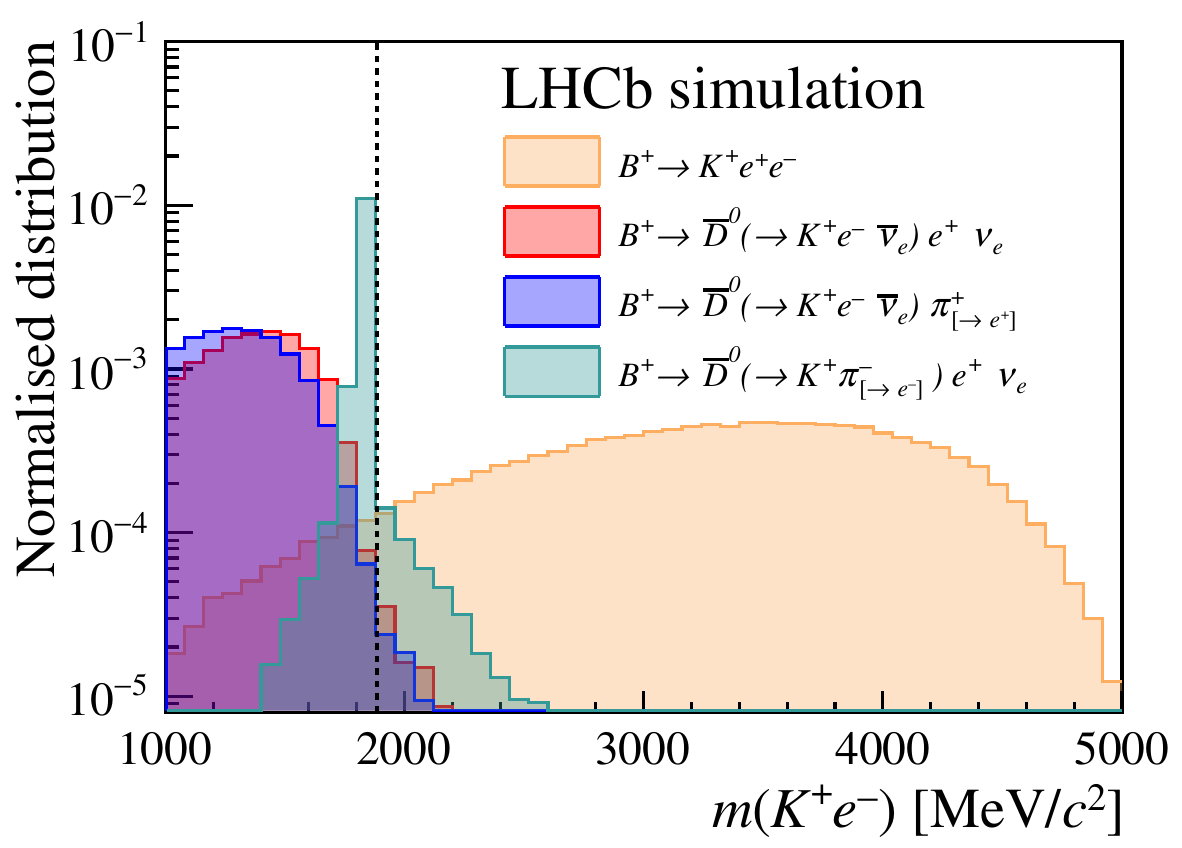}
      \includegraphics[width=0.48\linewidth]{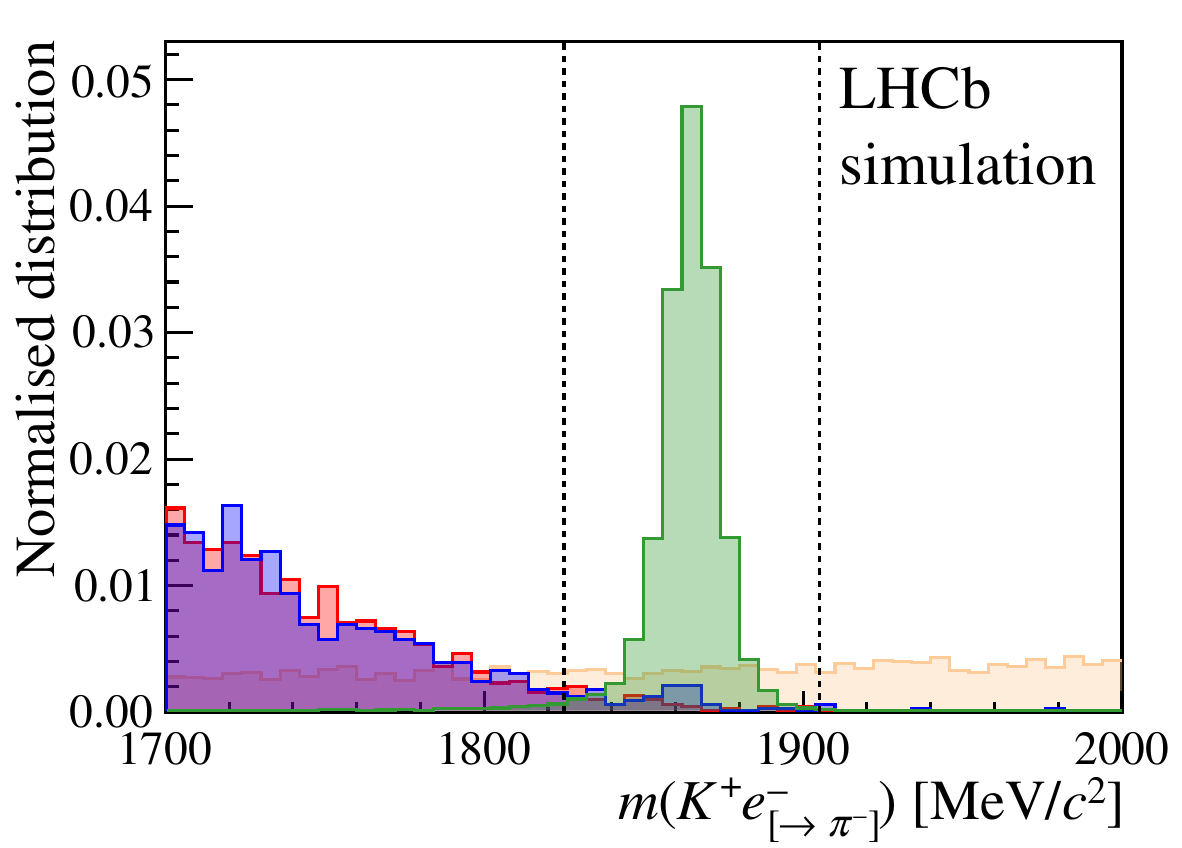}
   \end{center}
    \vspace*{-0.5cm}
   \caption{Simulated $K^+e^-$ mass distributions for  signal and various cascade background samples. \textcolor{changes}{The signal is represented by the orange shaded region and the various cascade background contributions by red, dark blue and light blue shaded regions.} The distributions are all normalised to unity.  (Left, with log $y$-scale)
   the bremsstrahlung correction to the momentum of the electron is applied, resulting in a tail to the right. The region to the left of the vertical dashed line is rejected. (Right, with linear $y$-scale) the mass is computed only from the track information. The notation $\pi^-_{[\rightarrow e^-]}$ ($e^-_{[\rightarrow \pi^-]}$) is used to denote an \textcolor{changes}{pion (electron)} that is \textcolor{changes}{reconstructed} as an \textcolor{changes}{electron (pion)}. The region between the dashed vertical lines is rejected. }\label{fig:cascadeVeto}
\end{figure}

Cascade background of the form \HbtoHc, where 
\Hc is a hadron containing a \cquark quark (\Dz, \Dp, \Ds, \Lc), and $X$, $Y$ are particles that are not included in the \Bu candidate, are suppressed by requiring that the kaon-lepton invariant mass is in the region \mbox{$m(\Kp\ellm)>m_{\Dz}$}, where $m_{\Dz}$ is the known $\Dz$ mass~\cite{PDG2020}. For the electron mode, this requirement is illustrated in Fig.~\ref{fig:cascadeVeto} (left). Analogous background sources with a misidentified particle are reduced by applying a similar veto, but with the lepton-mass hypothesis changed to that of a pion (denoted $\ell_{[\to\pi]}$). In the muon case, $K\mu_{[\to \pi]}$ combinations with a mass smaller than $m_{\Dz}$ are rejected. In the electron case, a $\pm40\mevcc$ window around the \Dz mass is used to reject candidates where the veto is applied without the bremsstrahlung recovery, \ie  based on only the measured track momenta. The mass distributions are shown in Fig.~\ref{fig:cascadeVeto}. The electron and muon veto cuts differ given the relative helicity suppression of $\pip\to\ell^+\nu_{\ell}$ decays. 
This causes misidentification backgrounds to populate a range of $K\mu$ masses but only a peak in the $Ke$ mass. The veto requirements retain 97\% of \BuKmm and 95\% of \BuKee decays passing all other selection requirements. 

Background from other exclusive $B$-hadron decays requires at least two particles to be misidentified. These include the decays \BuKpipi, and misreconstructed \BuJpsiKll and \BuPsiKll decays. In the latter two decays the kaon is misidentified as a lepton and the lepton (of the same electric charge) as a kaon. Such background is reduced to a negligible level by particle-identification criteria. Background from decays with a photon converted into an \epem pair are also negligible due to the \qsq selection.

\subsubsection*{Multivariate selection}

A Boosted Decision Tree (BDT)  algorithm~\cite{Breiman} with gradient boosting~\cite{GradBoost} is used to reduce combinatorial background. For the nonresonant muon mode and for each of the three different trigger categories of the 
nonresonant electron mode, a single BDT classifier is trained for the 7 and $8\tev$ data, and an additional classifier is trained for the $13\tev$ data. The BDT output is not strongly correlated with \qsq and the same classifiers are used to select the respective resonant decays. 
In order to train the classifier, simulated nonresonant \BuKll decays are used as a proxy for the signal and nonresonant \Kll candidates selected from the data with $\mKll > 5.4\gevcc$ are used as a background sample. The $k$-folding technique is used in the training and testing~\cite{Blum:1999:BHB:307400.307439}. 
The classifier includes the following variables:  the  \pt of the \Bp, \Kp and dilepton candidates, and the minimum and maximum \pt of the leptons;  the \Bp, dilepton and \Kp \chisqip with respect to the associated PV, where \chisqip is defined as the difference in the vertex-fit \chisq of the PV reconstructed with and without the considered particle; the minimum and maximum \chisqip of the leptons; the \Bp vertex-fit quality; the statistical significance of the \Bp flight distance; and the angle between the \Bp candidate momentum vector and the direction between the associated PV and the \Bp decay vertex. 
The \pt of the final state particles, the vertex-fit \chisq and the significance of the flight distance have the most discriminating power.
For each of the classifiers, a requirement is placed on the output variable in order to maximise the predicted significance of the nonresonant signal yield. For the electron modes that dictate the \RK precision, this requirement reduces the combinatorial background by approximately $99\%$, while retaining $85\%$ of the signal mode. The muon BDT classifier has similar performance. In both cases, for both signal and background, the efficiency of the BDT selection has negligible dependence on \mKll and \qsq in the regions used to determine the event yields. 

\subsubsection*{Calibration of simulation} 

The simulated data used in this analysis are produced using the software described in Refs.~\cite{Sjostrand:2006za,*Sjostrand:2007gs,LHCb-PROC-2010-056,Lange:2001uf,Allison:2006ve, *Agostinelli:2002hh,LHCb-PROC-2011-006}. Bremsstrahlung emission in the decay of particles is simulated using the \photosplusplus software in the default configuration~\cite{ Davidson:2010ew}, which is observed to agree with an independent quantum electrodynamics calculation at the level of $1\%$~\cite{Bordone:2016gaq}.

Simulated events are weighted to correct for the imperfect modelling using control channels. The \Bu production kinematics are corrected using \BuJpsiKll events. The particle-identification performance is calibrated using data, where the species of particles in the final state can be unambiguously determined purely on the basis of the kinematics. 
The calibration samples consist of $\decay{\Dstarp}{\Dz(\to\Km\pip)\pip}$, $\decay{\jpsi}{\mumu}$, and \BuJpsiKee decays, from which kaons, muons, and electrons, respectively, can be selected without applying particle-identification requirements. The performance of the particle-identification requirements is then evaluated from the proportion of events in these samples which fulfil the particle-identification selection criteria. The trigger response is corrected using weights applied to simulation as a function of variables relevant to the trigger algorithms. The weights are calculated by requiring that simulated \BuJpsiKll events  exhibit the same trigger performance as the control data. The \BuJpsiKll events selected from the data have also been used to demonstrate control of the electron track-reconstruction efficiency at the percent level~\cite{Aaij:2019vvl}.
Whenever \BuJpsiKll events are used to correct the simulation, the correlations between calibration and measurement samples are taken into account in the results and cross-checks presented in the article. 
The correlation is evaluated using a bootstrapping method to recompute the yields and efficiencies many times with different subsets of the data.

\subsubsection*{Likelihood fit} 

An unbinned extended maximum-likelihood fit is made to the \mKee and \mKmm distributions of nonresonant candidates. The value of \RK is a fit parameter, which is related to the signal yields and efficiencies according to  
\begin{eqnarray}
    \label{eq:RKdetail}
       \RK &=&\frac{N(\BuKmm)}{\varepsilon(\BuKmm)}\cdot\frac{\varepsilon(\BuKee)}{N(\BuKee)} \times \nonumber\\
 & & \frac{\varepsilon(\BuJpsiKmm)}{N(\BuJpsiKmm)}\cdot\frac{N(\BuJpsiKee)}{\varepsilon(\BuJpsiKee)} \,, 
\end{eqnarray}
\noindent where $N(X)$  indicates the  yield  of  decay  mode $X$  
and $\varepsilon(X)$ is the efficiency for selecting decay mode $X$. 
The resonant yields are determined from separate fits to \mKllconst. 
In the fit for \RK these yields, together with the efficiencies, are incorporated as Gaussian-constraint terms.

In order to take into account the correlation between the selection efficiencies, the \mKee and \mKmm distributions of nonresonant candidates in each of the different trigger categories and data-taking periods are fitted simultaneously, with a common value of \RK. The relative fraction of partially reconstructed background in each trigger category is also shared across the different data-taking periods.

The mass-shape parameters are derived from the calibrated simulation. The four signal modes are modelled by multiple Gaussian functions with  power-law tails on both sides of the peak~\cite{Skwarnicki:1986xj,Santos:2013gra}, although the differing detector response gives different shapes for the electron and muon modes. The signal mass shapes of the electron modes are described with the sum of three distributions, which model whether the ECAL energy deposit from a bremsstrahlung photon was added to both, either, or neither of the \epm candidates. The expected values from simulated events are used to constrain the fraction of signal decays in each of these categories. These fractions are observed to agree well with those observed in resonant events selected from the data. 
In order to take into account residual differences in the signal shape between data and simulation, an offset in the peak position and a scaling of the resolution are allowed to vary in the fits to the resonant modes. The corresponding parameters are then fixed in the fits to the relevant nonresonant modes. \textcolor{changes}{This resolution scaling changes the migration of candidates into the \qsq region of interest by less than 1\%.}

For the modelling of nonresonant and resonant partially reconstructed backgrounds, data are used to correct the simulated $K\pi$ mass spectrum for \BuBdKpiplusee and \BuBdKpijpsi decays~\cite{LHCb-PAPER-2016-025}. The calibrated simulation is used subsequently to obtain the \mKll mass shape and relative fractions of these background components. In order to accommodate possible lepton-universality violation in these partially reconstructed processes, which are underpinned by the same $\bquarkbar\to\squarkbar$ quark-level transitions as those of interest, the overall yield of such decays is left to vary freely in the fit. The shape of the \BuJpsipi background contribution is taken from simulation but the size with respect to the \BuJpsiK mode is constrained using the known ratio of the relevant branching fractions~\cite{LHCb-PAPER-2016-051, PDG2020} and efficiencies.

In the fits to nonresonant \BuKee candidates, the mass shape of the background from \BuJpsiKee decays with an emitted photon that is not reconstructed is also taken from simulation and, adjusting for the relevant selection efficiency, its yield is constrained to the value from the fit to the resonant mode within its uncertainty.   
In all fits, the combinatorial background is modelled with an exponential function with a freely varying yield and shape.

\begin{figure}
    \centering
    \includegraphics[width=0.45\textwidth]{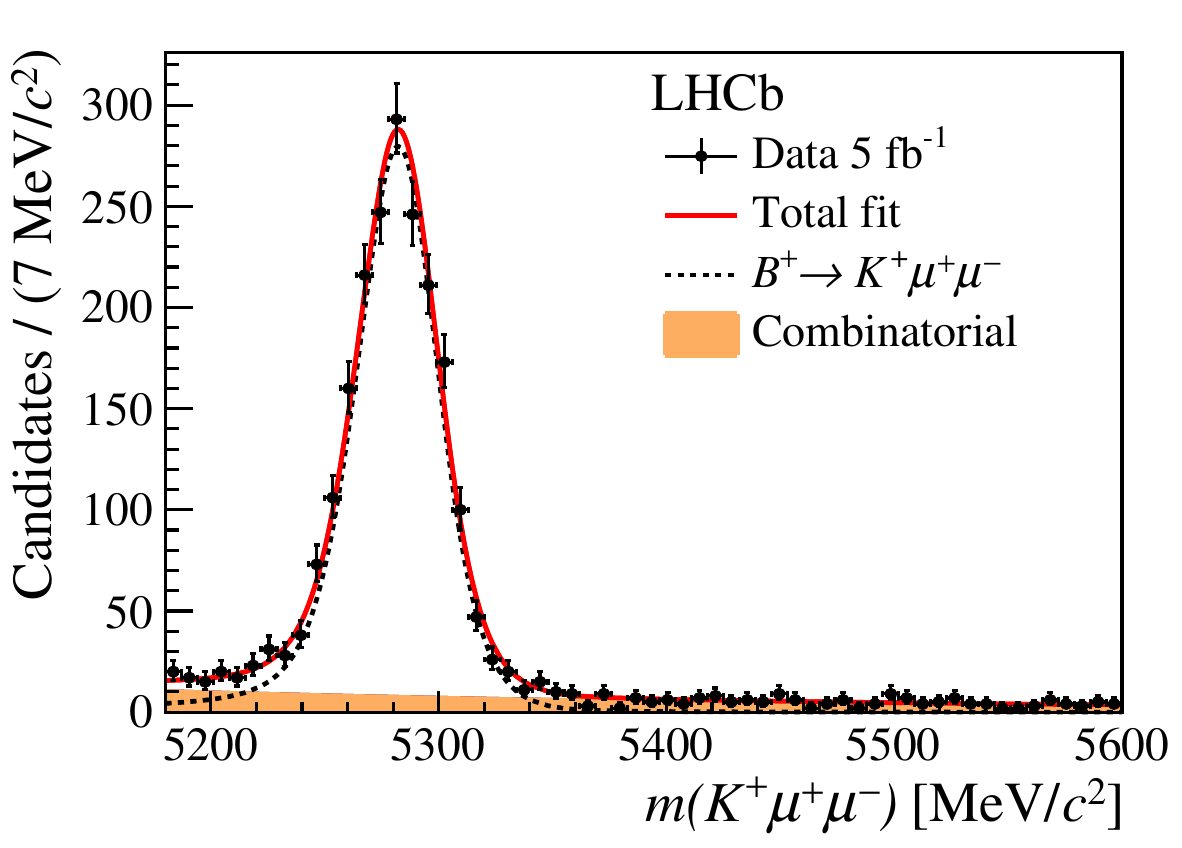}
    \includegraphics[width=0.45\textwidth]{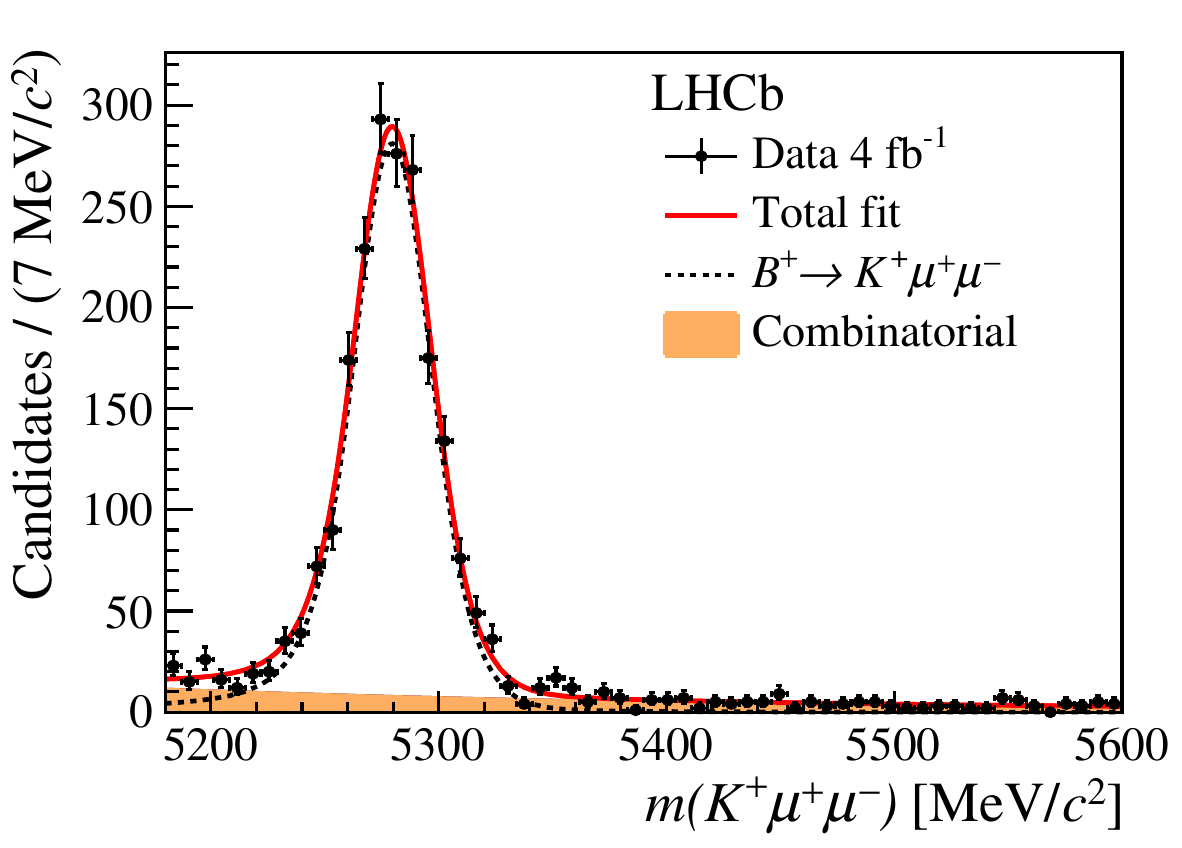}
                                                      
    \includegraphics[width=0.45\textwidth]{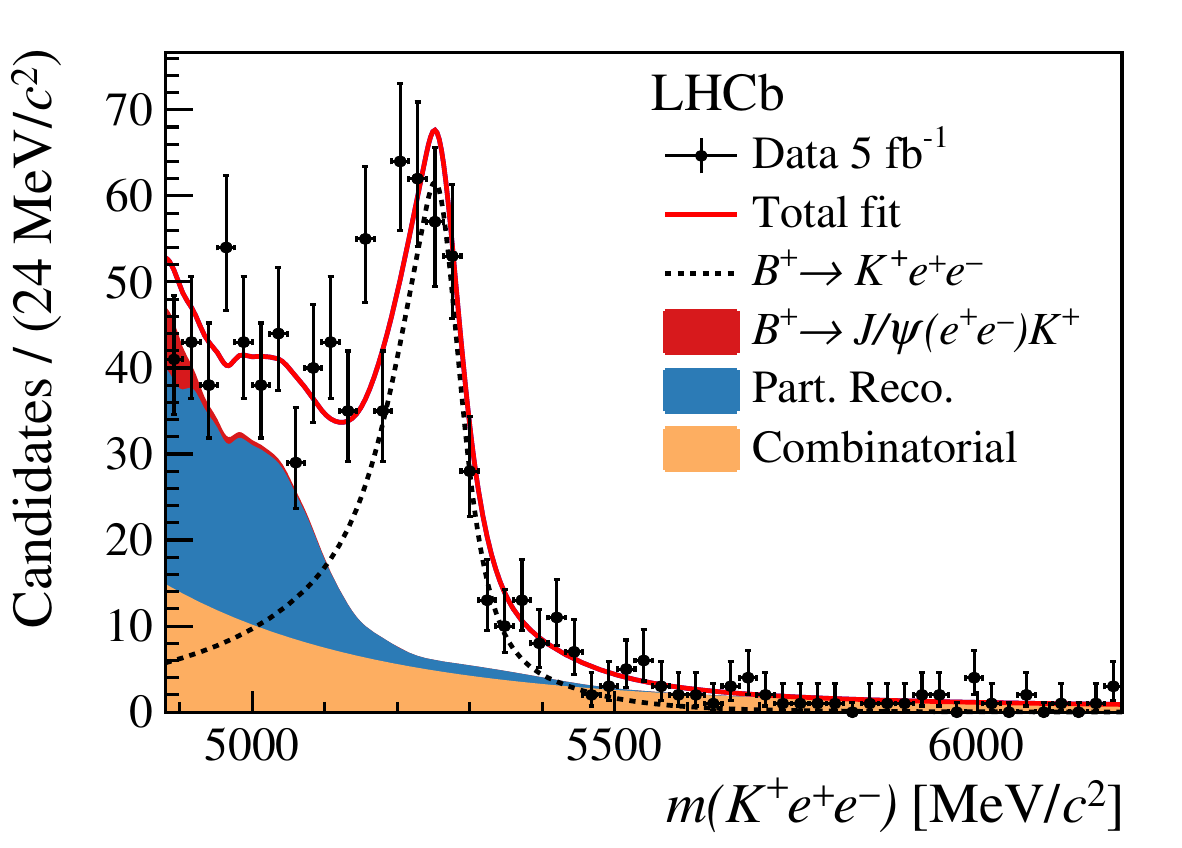}
    \includegraphics[width=0.45\textwidth]{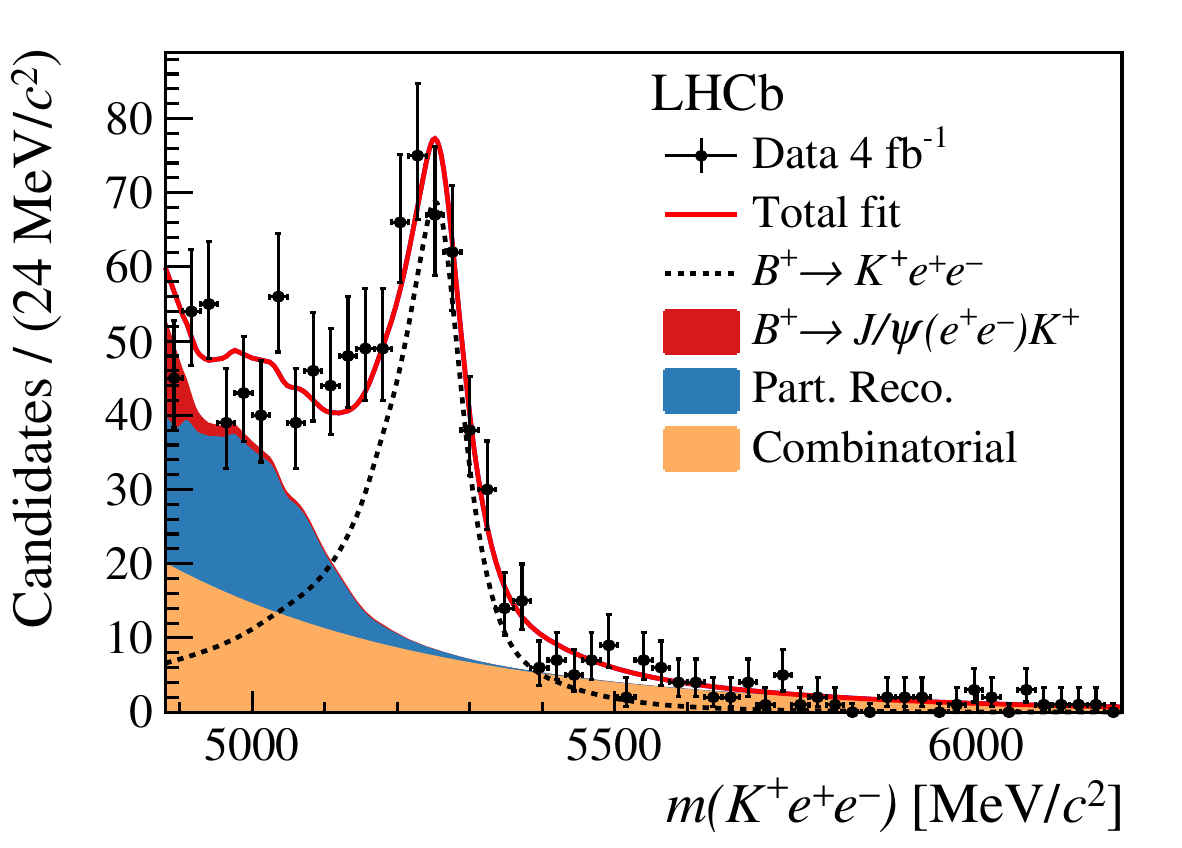}
                                                      
    \includegraphics[width=0.45\textwidth]{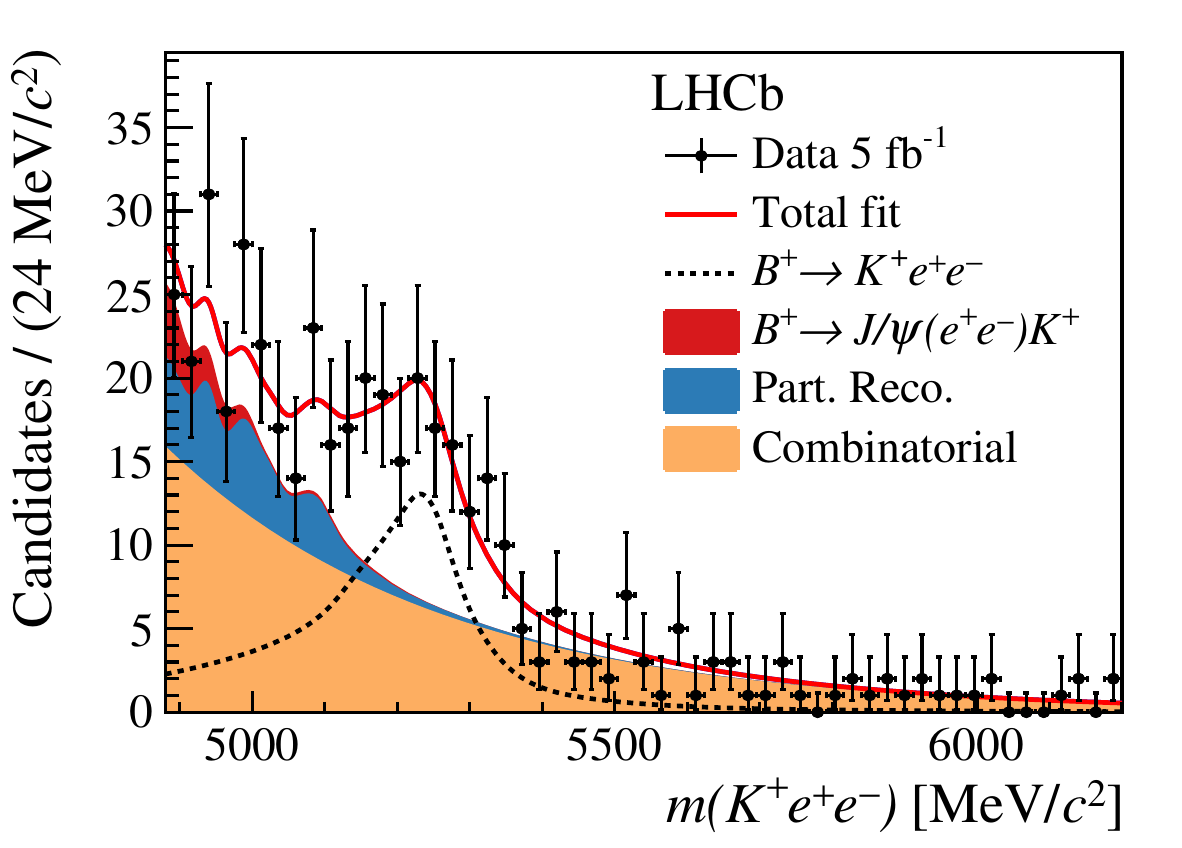}
    \includegraphics[width=0.45\textwidth]{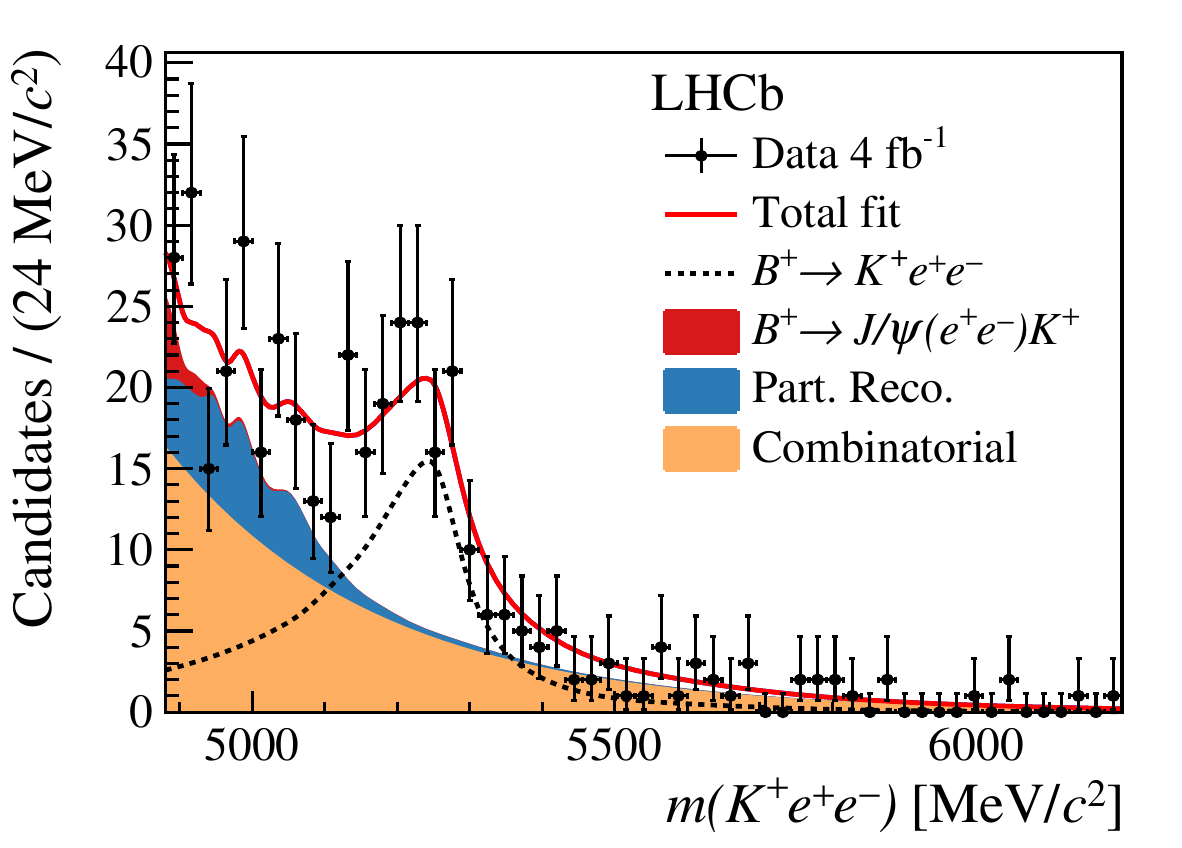}
                                                      
    \includegraphics[width=0.45\textwidth]{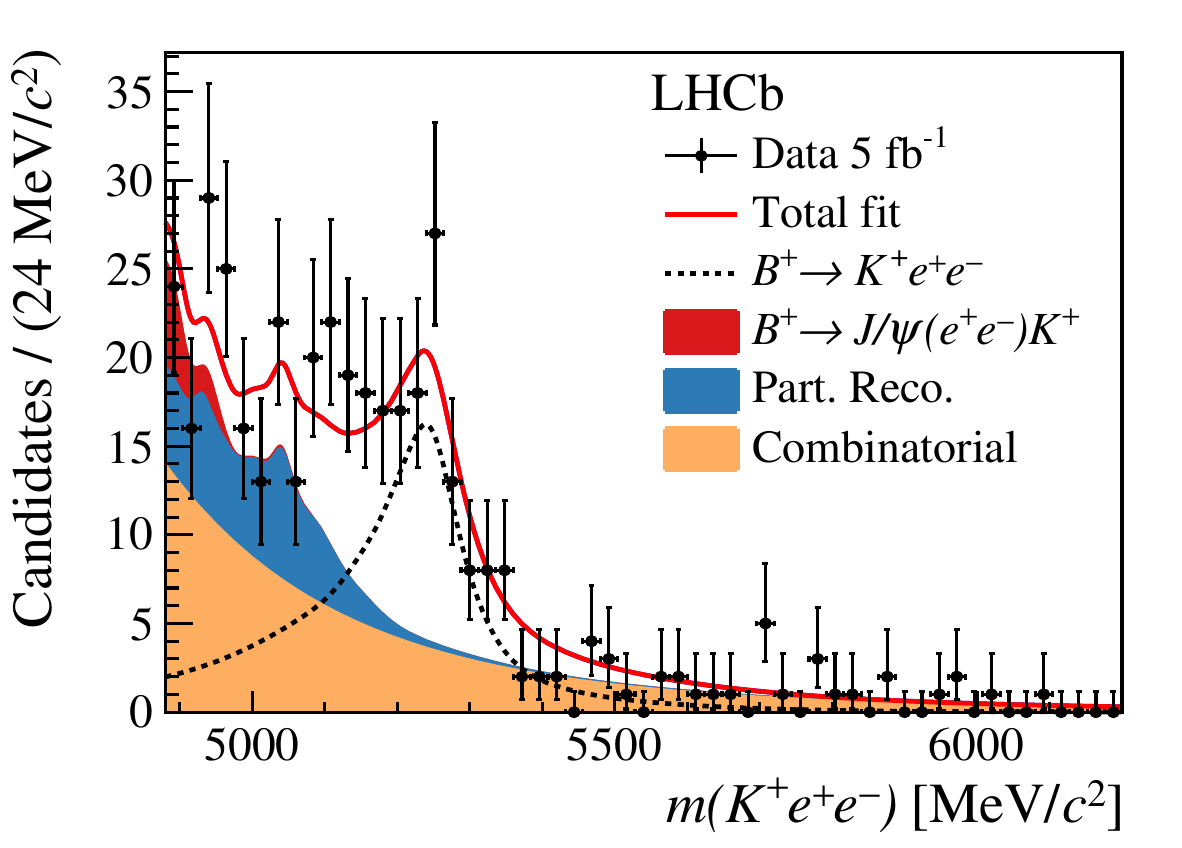}
    \includegraphics[width=0.45\textwidth]{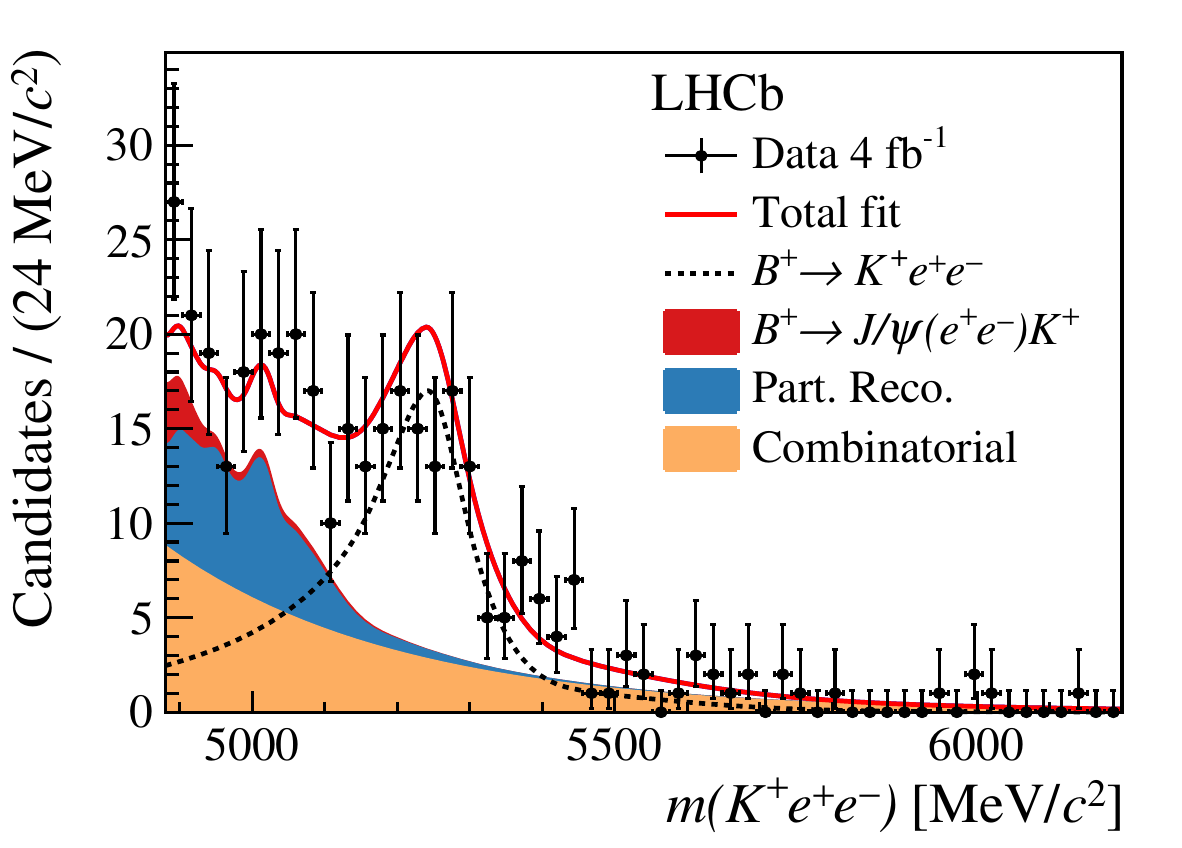}
    \caption{Candidate invariant mass distributions. Distribution of the invariant mass \mKll for nonresonant candidates in the (left) sample previously analysed~\cite{LHCb-PAPER-2019-009} and (right) the new data sample. The top row shows the fit to the muon modes and the subsequent rows the fits to the electron modes triggered by (second row) one of the electrons, (third row) the kaon and (last row) by other particles in the event. The fit projections are superimposed\textcolor{changes}{, with dotted lines describing the signal contribution and solid areas representing each of the background components described in the text and listed in the legend}.
    }
    \label{fig:nonresfits_categories}
\end{figure}

\begin{figure}
    \centering
    \includegraphics[width=0.45\textwidth]{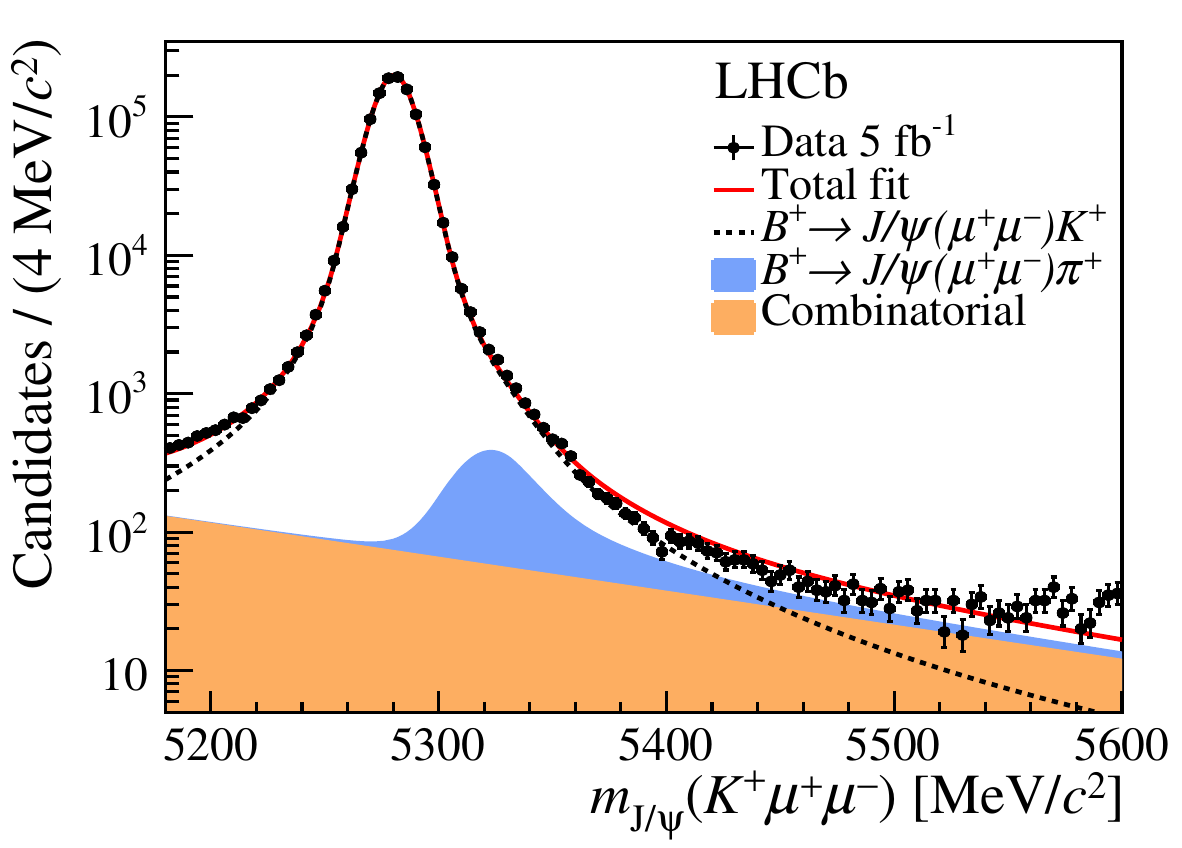}
    \includegraphics[width=0.45\textwidth]{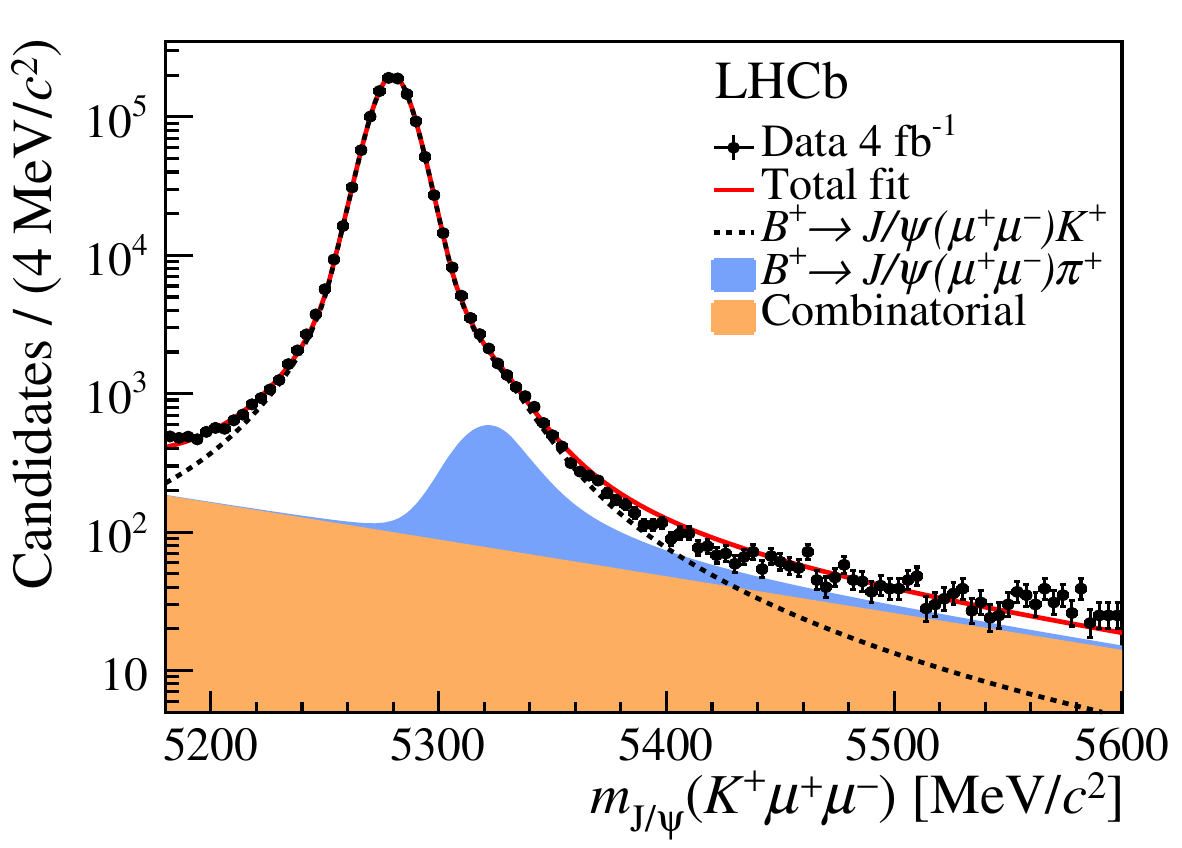}
                                                      
    \includegraphics[width=0.45\textwidth]{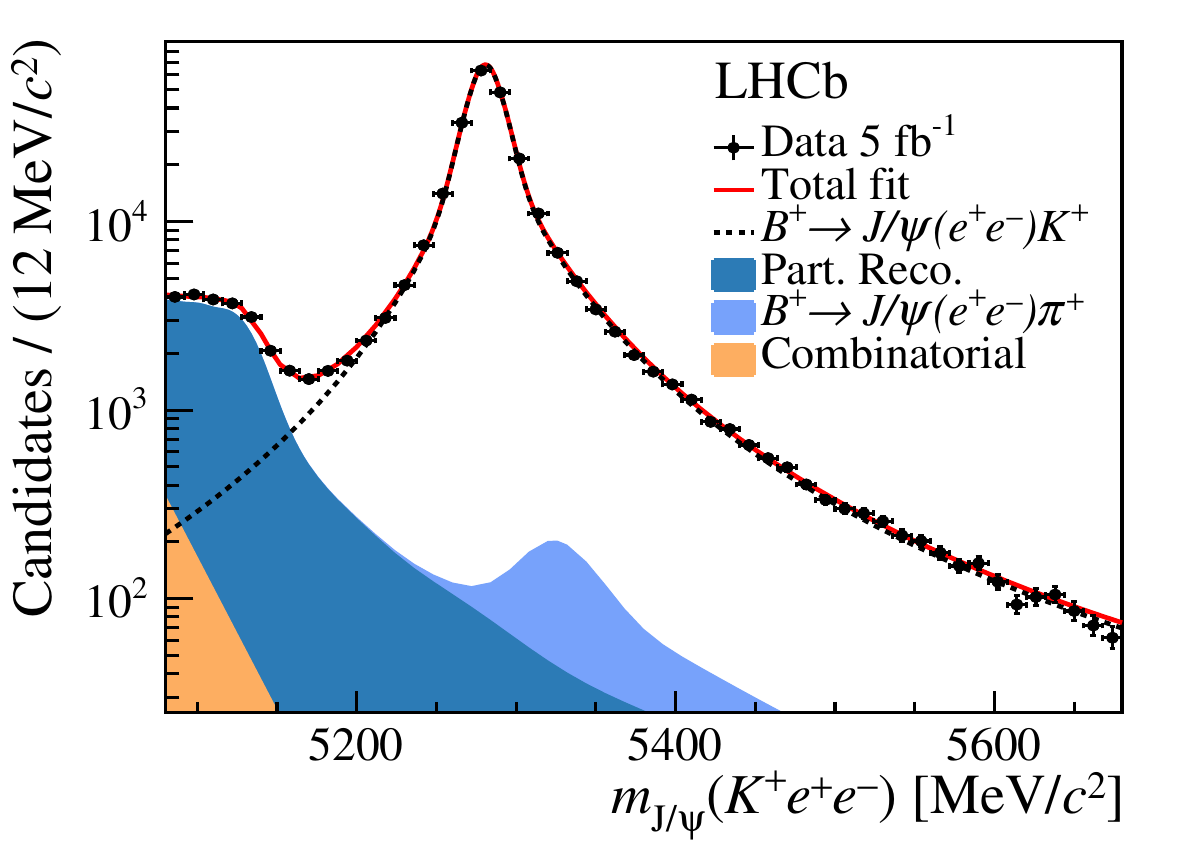}
    \includegraphics[width=0.45\textwidth]{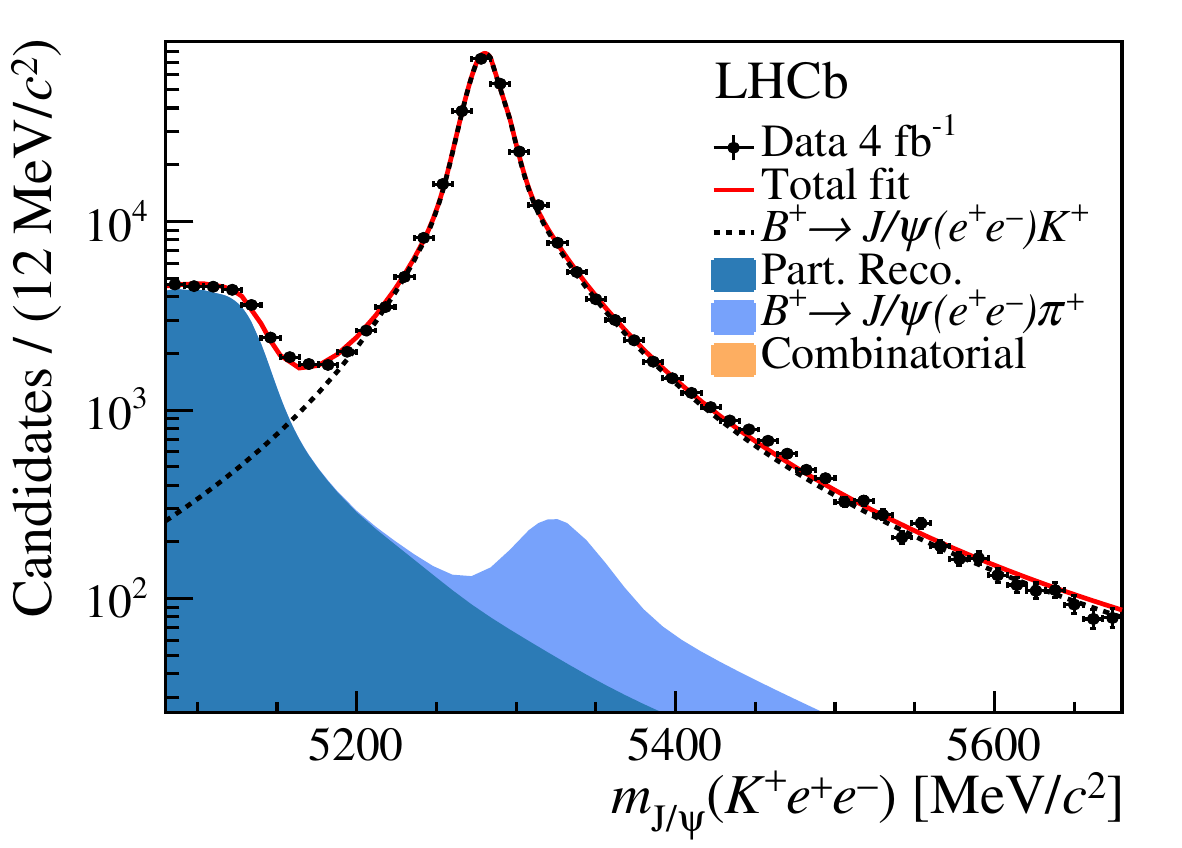}
                                                      
    \includegraphics[width=0.45\textwidth]{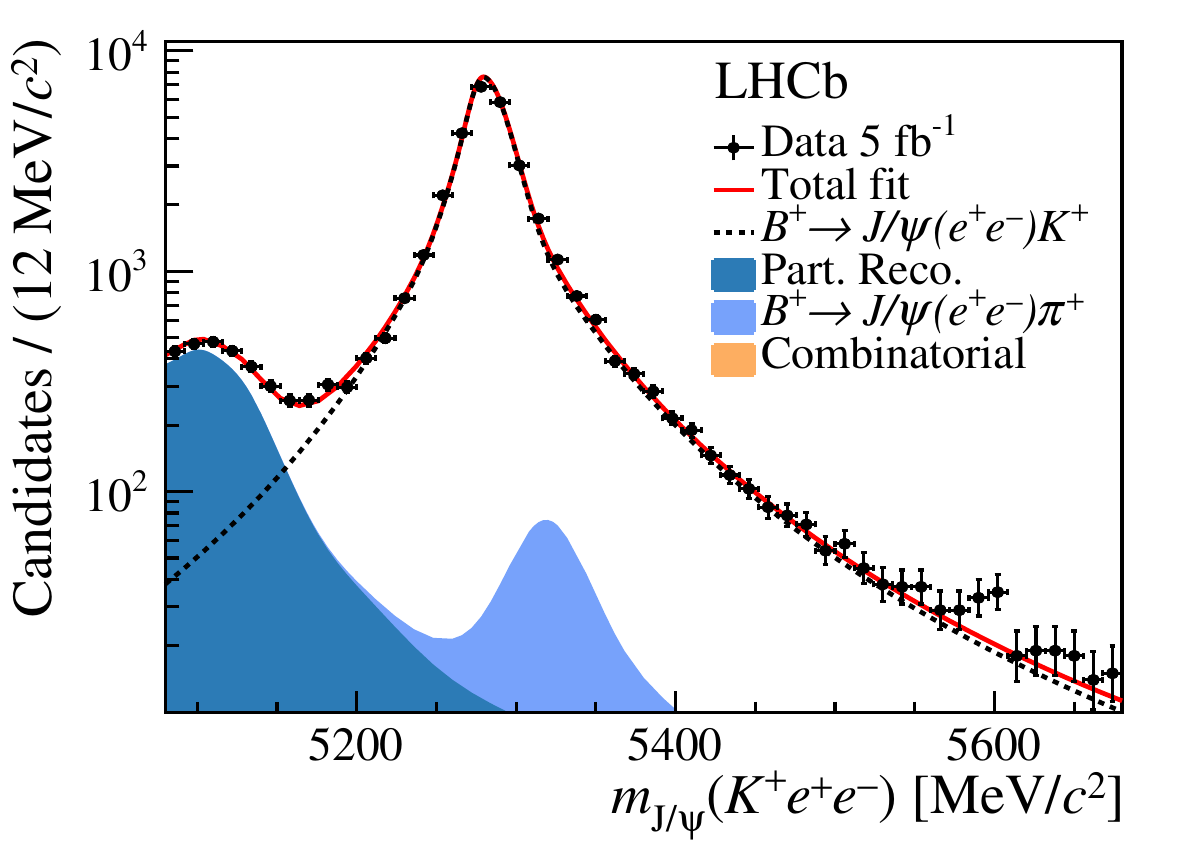}
    \includegraphics[width=0.45\textwidth]{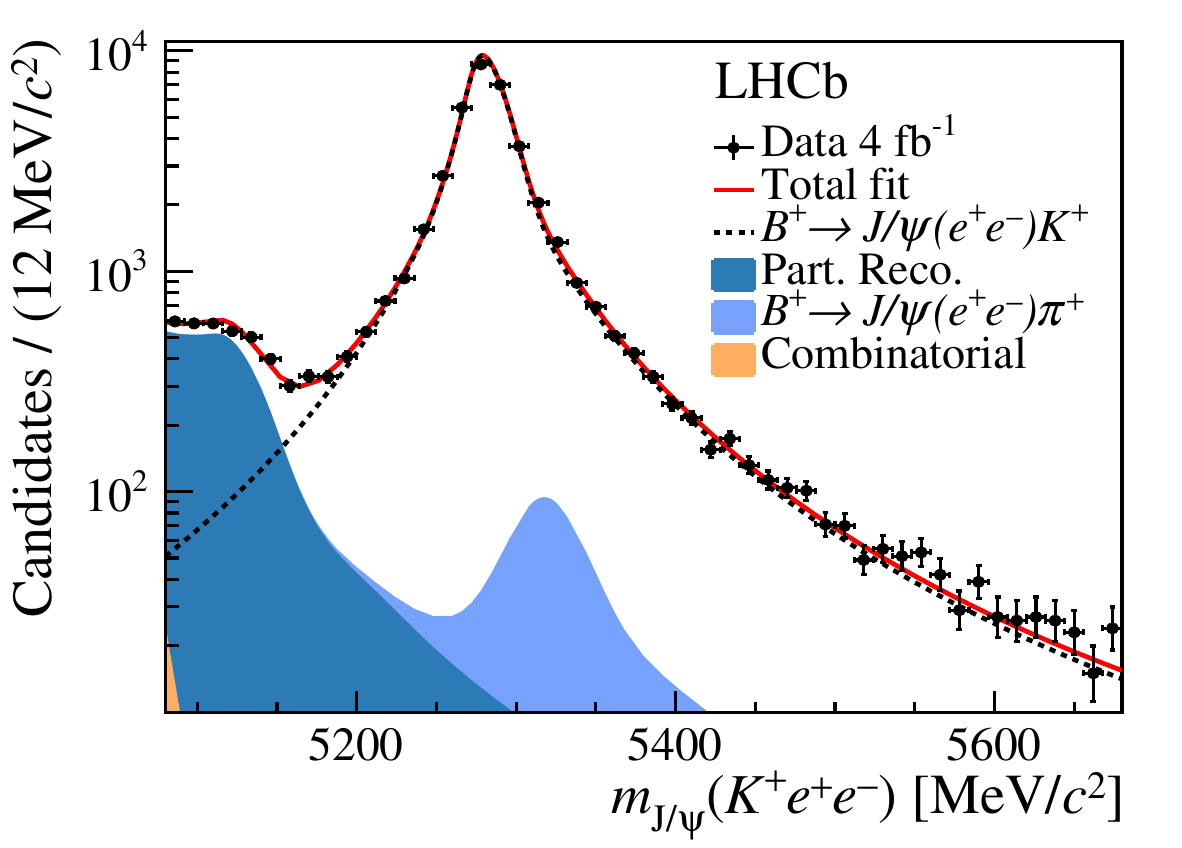}
                                                      
    \includegraphics[width=0.45\textwidth]{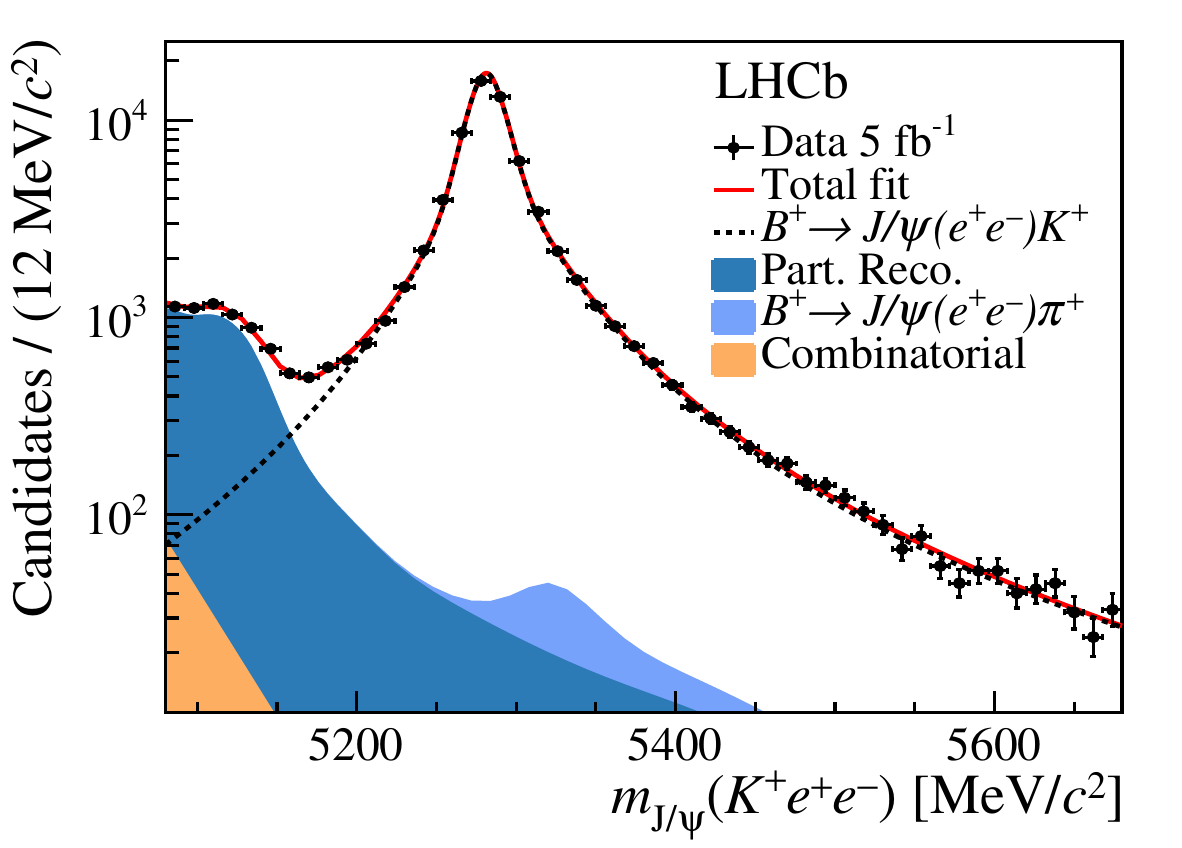}
    \includegraphics[width=0.45\textwidth]{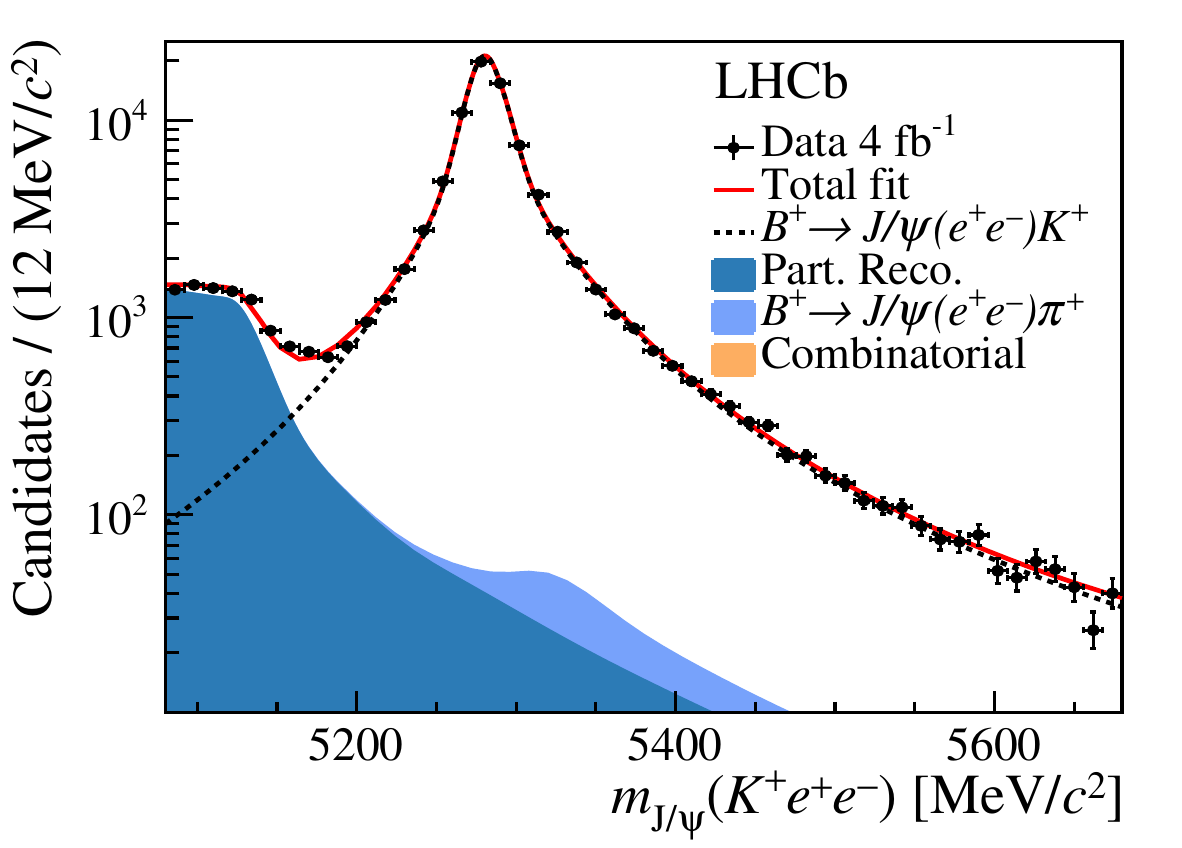}
    
    \caption{Candidate invariant mass distributions. Distribution of the invariant mass \mKllconst for resonant candidates in the (left) sample previously analysed~\cite{LHCb-PAPER-2019-009} and (right) the new data sample. The top row shows the fit to the muon modes and the subsequent rows the fits to the electron modes triggered by (second row) one of the electrons, (third row) the kaon and (last row) by other particles in the event. The fit projections are superimposed\textcolor{changes}{, with dotted lines describing the signal contribution and solid areas representing each of the background components described in the text and listed in the legend}.
}
    \label{fig:resfits_categories}
\end{figure}

The fits to the nonresonant (resonant) decay modes in different data-taking periods and trigger categories are shown in Fig.~\ref{fig:nonresfits_categories}  (Fig.~\ref{fig:resfits_categories}).
For the resonant modes the results from independent fits to each period/category are shown. Conversely, the nonresonant distributions show the projections from the simultaneous fit across data taking periods and trigger categories that is used to obtain \RK. The fitted yields for the resonant and nonresonant decays are given in Table~\ref{tab:yields}. 

\begin{table}[b]
\centering
\caption{Yields of the  nonresonant and resonant decay modes obtained from the fits to the data. The quoted uncertainty is the combination of statistical and systematic effects.} \label{tab:yields}
\begin{tabular}{lc}
\toprule
Decay mode  &   Yield             \\
\midrule
\BuKee      &   $\phantom{0\,00}1\,640 \pm \phantom{0\,0}70$ \\
\BuKmm      &   $\phantom{0\,00}3\,850 \pm \phantom{0\,0}70$ \\
\BuJpsiKee  &   $\phantom{0\,}743\,300 \pm \phantom{0\,}900$ \\
\BuJpsiKmm  &   $2\,288\,500 \pm 1\,500$      \\
\bottomrule
\end{tabular}
\end{table}

\begin{figure}[!t]
   \begin{center}
      \includegraphics[height=0.25\textheight]{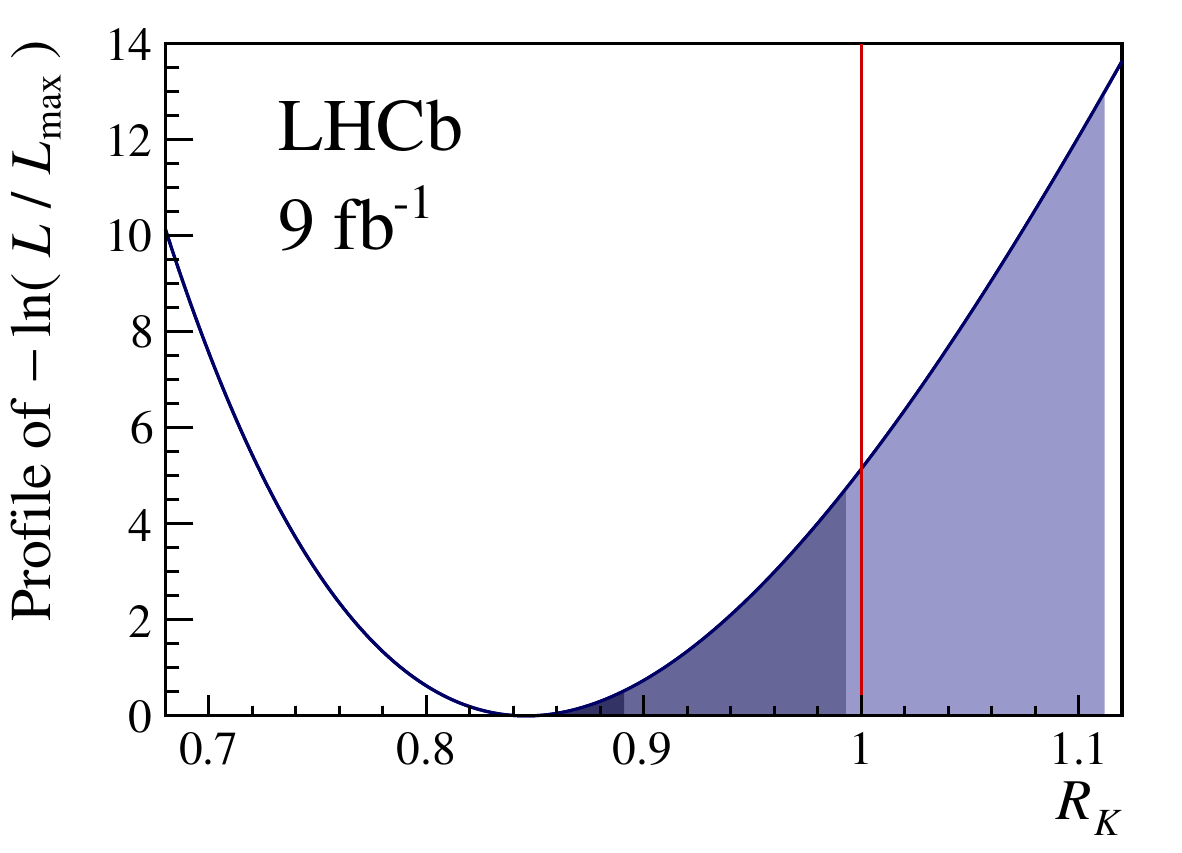}
  \end{center}
     \caption{\textcolor{changes}{Likelihood function from the fit to the nonresonant \BuKll candidates in terms of the ratio between the likelihood value ($L$) and that found by the fit ($L_{\rm max}$) as a function of \RK.} The extent of the  dark, medium and light blue  regions shows the values allowed for \RK at $1\sigma$, $3\sigma$ and $5\sigma$ levels. The red line indicates the prediction from the SM. }
    \label{fig:profile_likelihood}
\end{figure}

The profile likelihood for the fit to the nonresonant decays is shown in Fig.~\ref{fig:profile_likelihood}. The likelihood is non-Gaussian in the region $\RK>0.95$ due to the comparatively low yield of \BuKee events. 
Following the procedure described in Refs.~\cite{LHCb-PAPER-2019-009, LHCb-PAPER-2017-013}, the p-value is computed by integrating the posterior probability density function for \RK, having folded in the theory uncertainty on the SM prediction, for \RK values larger than the SM  expectation. The corresponding significance in terms of standard deviations is computed using the inverse Gaussian cumulative distribution function for a one-sided conversion.

A test statistic is constructed that is based on the likelihood ratio between two hypotheses with common (null) or different (test) \RK values for the part of the sample analysed previously (7, 8 and part of the 13\tev data) and for the new portion of the 13\tev data. Using pseudoexperiments based on the null hypothesis, the data suggest that the \RK value from the new portion of the data is compatible with that from the previous sample with a p-value of 95\%. Further tests give good compatibility for subsamples of the data corresponding to different trigger categories and magnet polarities.

The departure of the profile likelihood shown in Fig.~\ref{fig:profile_likelihood} from a normal distribution stems from the  definition of \RK. In particular,
in the \RK ratio the denominator is affected by larger statistical uncertainties than the numerator, owing to the larger number of nonresonant muonic signal candidates. However, the intervals of the likelihood distribution are found to be the same when estimated with $1/\RK$ as the fit parameter.

\subsubsection*{Additional cross-checks} 

\begin{figure}[!t]
   \begin{center}
   \begin{overpic}[width=0.45\linewidth,trim={0 0 0 0.5cm}, clip]{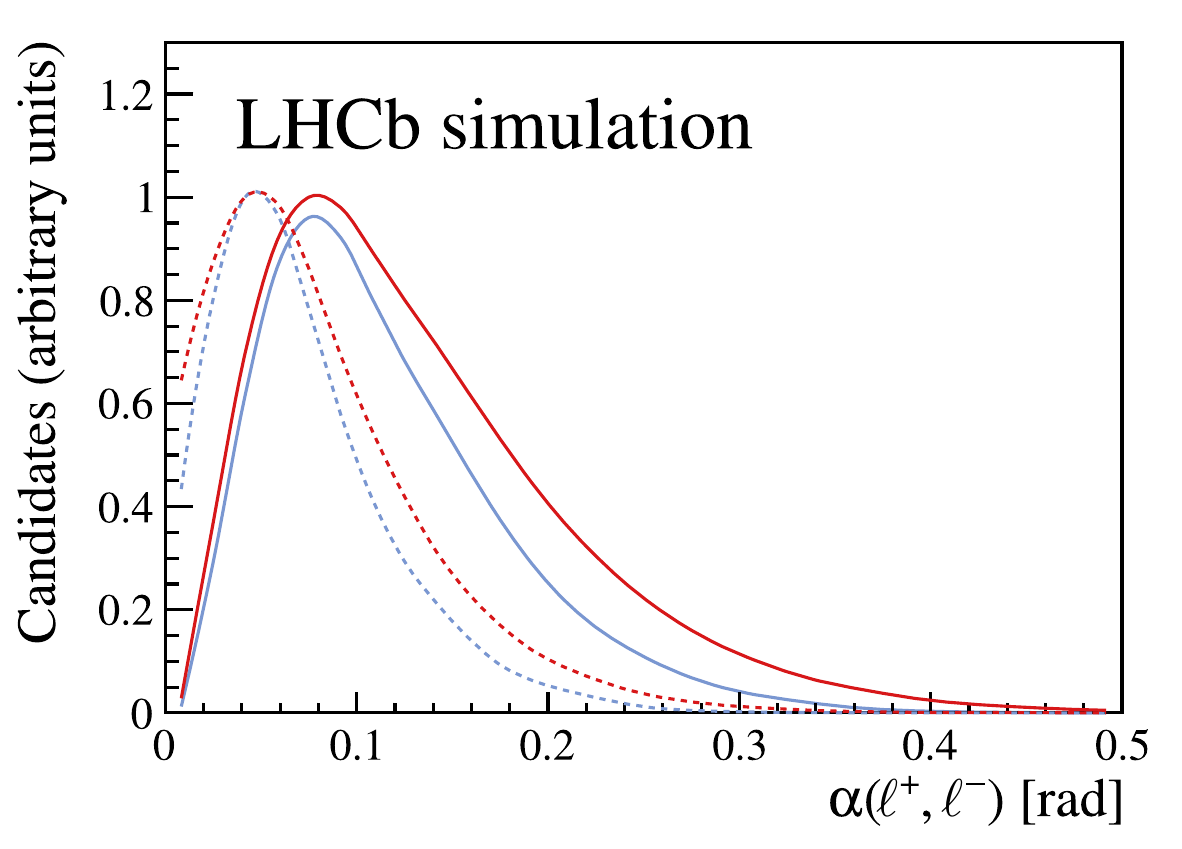}
   \put(50,26){\includegraphics[width=0.2\linewidth]{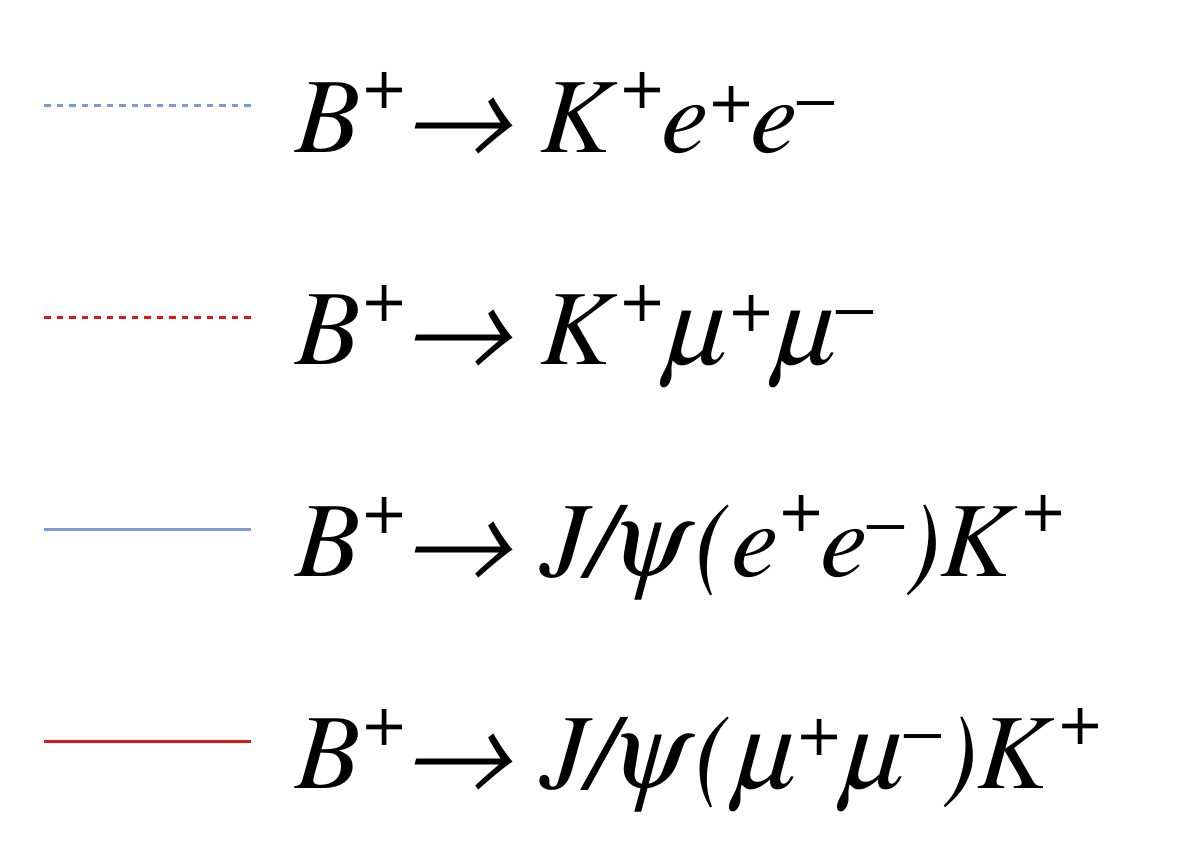}}
   \end{overpic}
   \includegraphics[width=0.45\linewidth,trim={0 0.15cm 0 0}, clip]{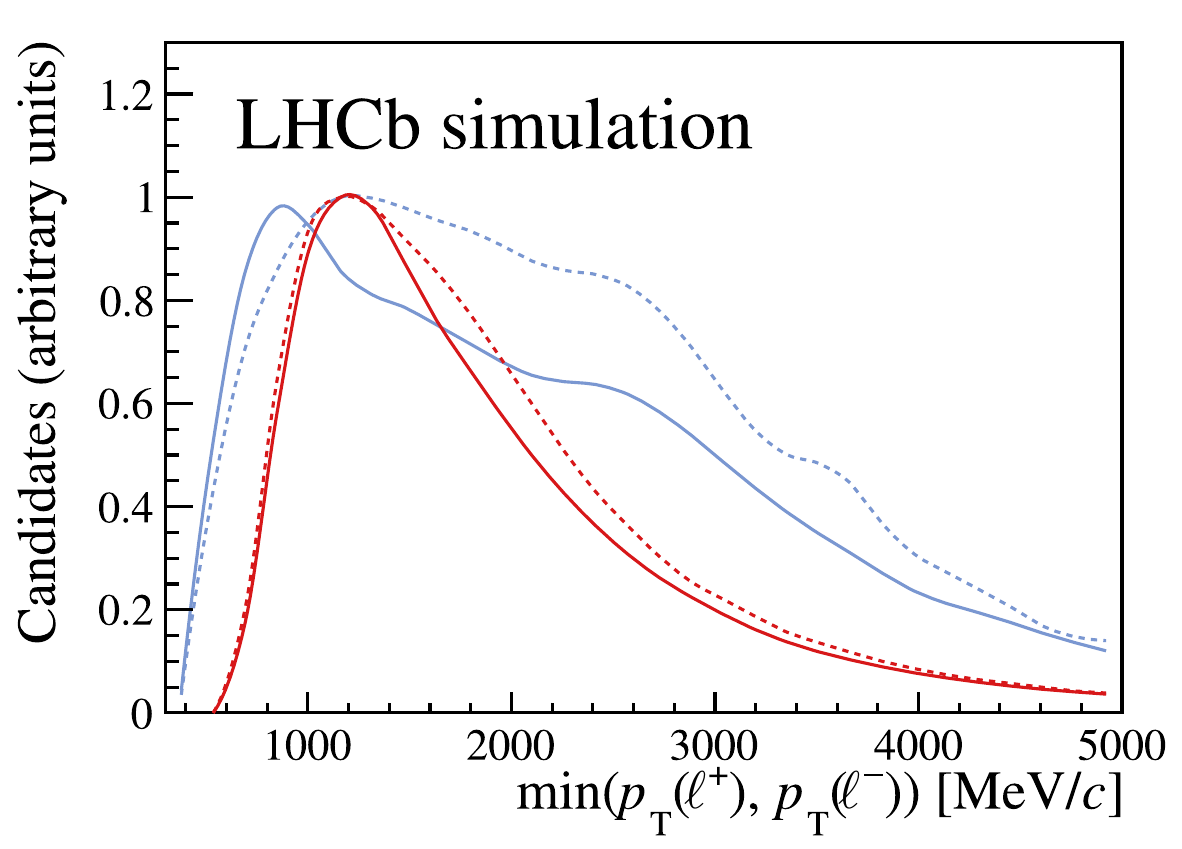}
  
  \includegraphics[width=0.45\linewidth,trim={0 0 0 0.5cm}, clip]{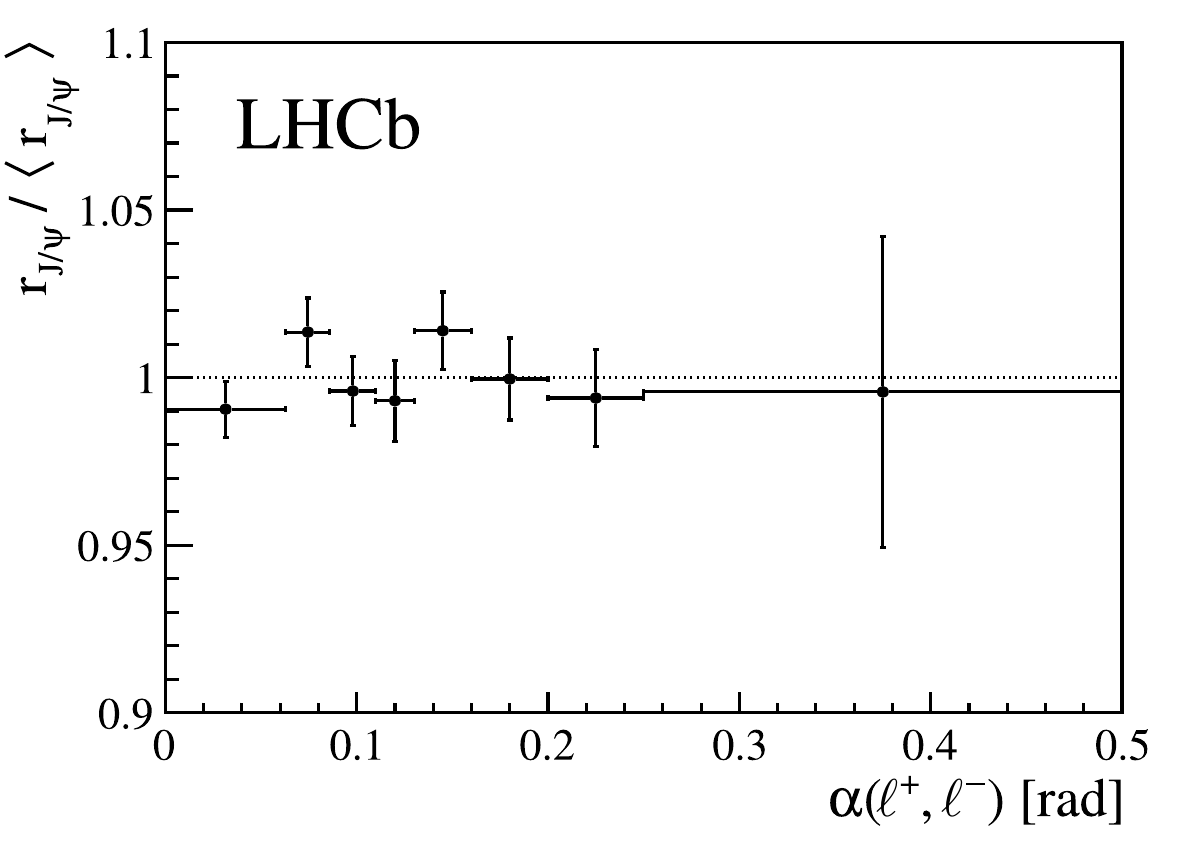}
   \includegraphics[width=0.45\linewidth,trim={0 0 0 0.5cm}, clip]{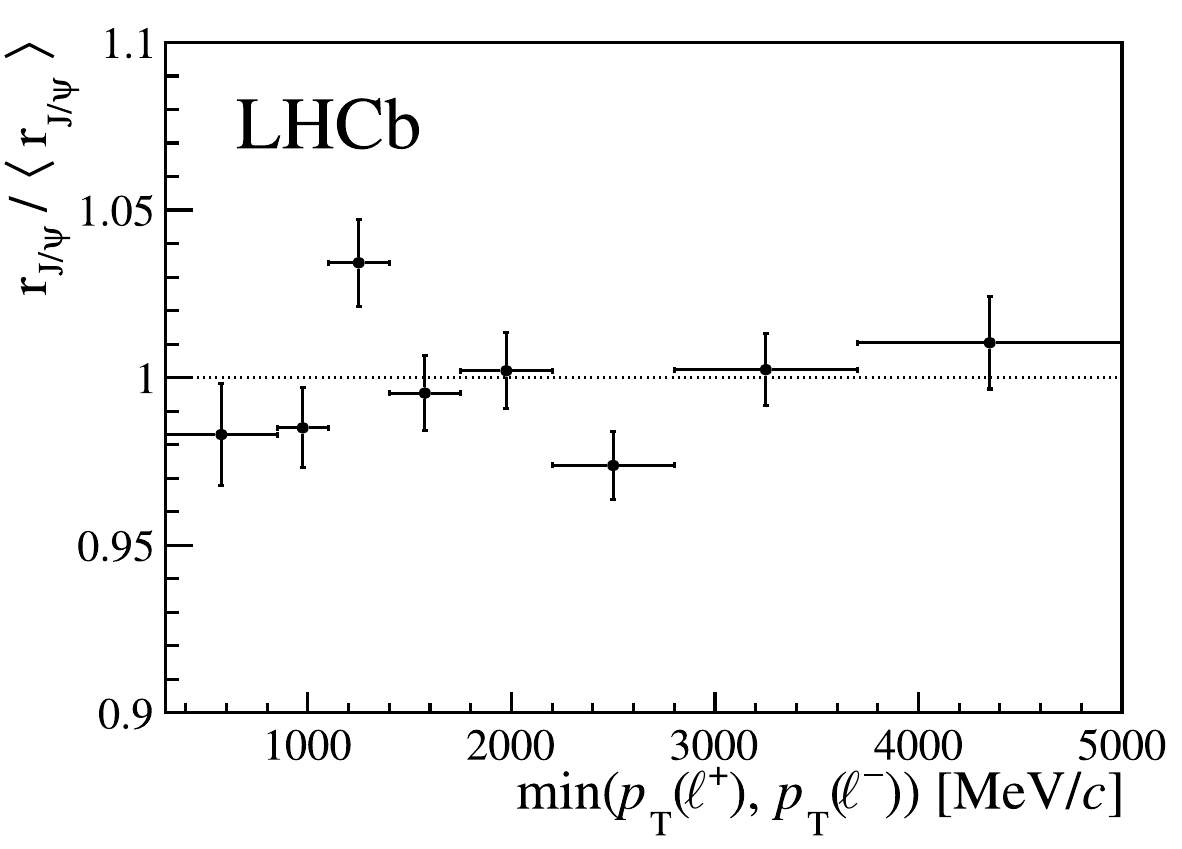}
      \end{center}
     \caption{Differential \rjpsi measurement. (Top) distributions of the reconstructed spectra of (left) the angle between the leptons\textcolor{changes}{, $\alpha(\ell^+, \ell^-)$}, and (right) the minimum \pt of the leptons \textcolor{changes}{ for \BuKll and \BuJpsiKll decays}. (Bottom) the single ratio \rjpsi relative to its average value $\left< \rjpsi \right>$ as a function of these variables. In the electron minimum \pt spectra, the structure at 2800\mevc is related to the trigger threshold. \textcolor{changes}{ Uncertainties on the data points are statistical only.}}
    \label{fig:rjpsi_differential1}
\end{figure}

\begin{figure}[!htbp]
   \begin{center}
   \includegraphics[width=0.45\linewidth]{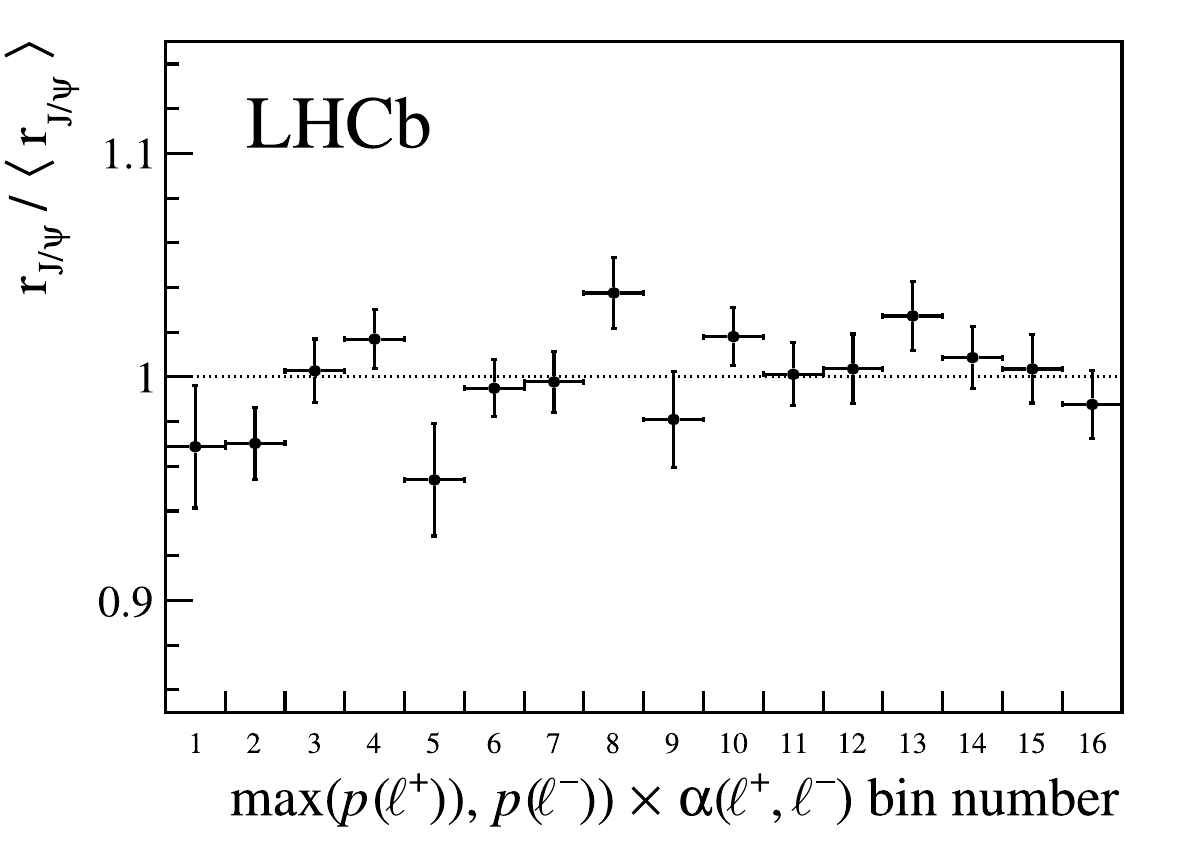}
    \includegraphics[height=0.32\linewidth]{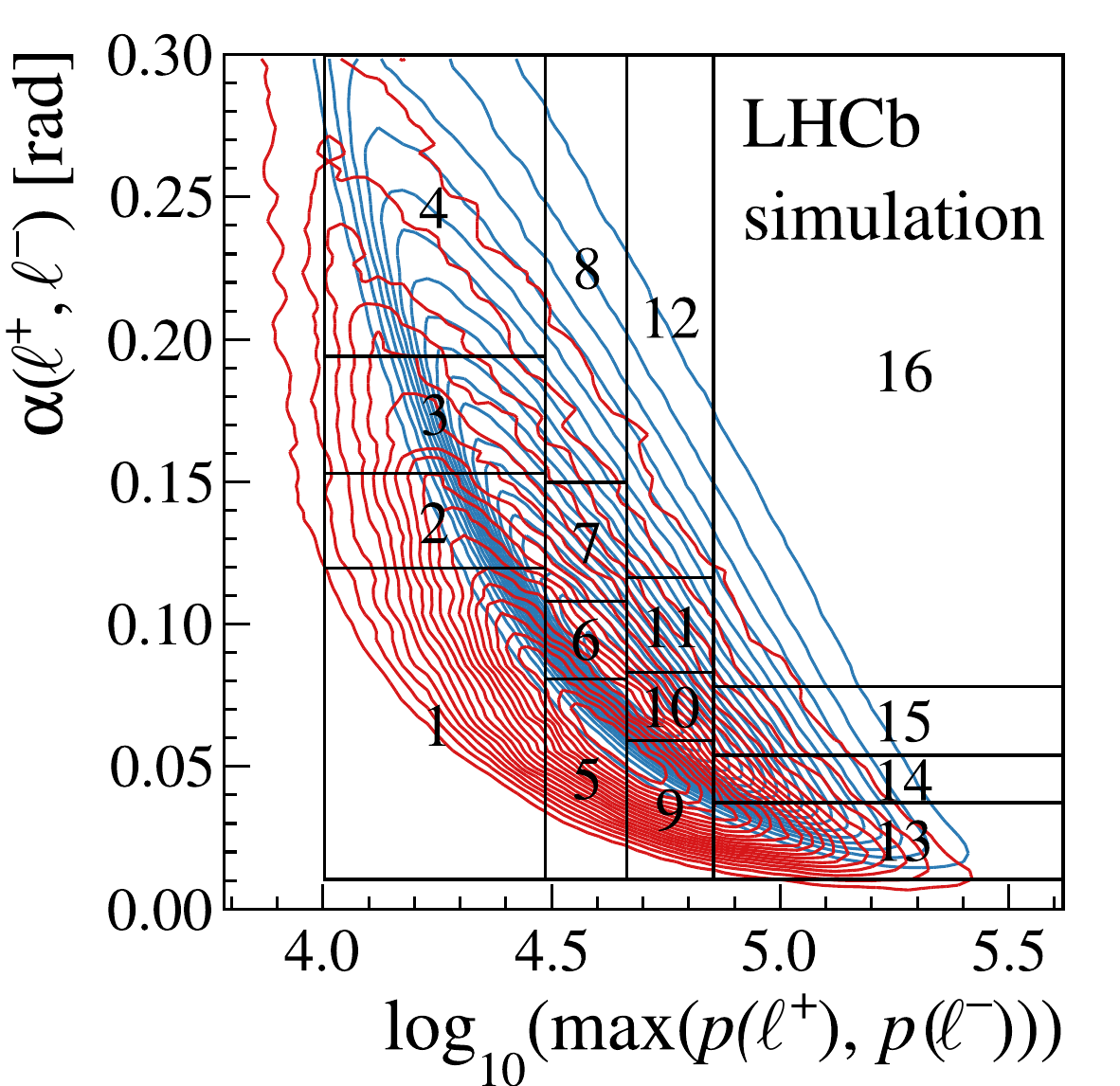}
   \end{center}
     \caption{Double differential \rjpsi measurement. (Left) the value of \rjpsi, relative to the average value of \rjpsi, measured in two-dimensional bins of the maximum lepton momentum, $p(\ell)$, and the opening angle between the two leptons, $\alpha(\ell^+,\ell^-)$. (Right) the bin definition in this two-dimensional space together with the
     distribution for \BuKee (\BuJpsiKee) decays depicted as red (blue) contours.  \textcolor{changes}{Uncertainties on the data points are statistical only.}}
    \label{fig:rjpsi_bin}
\end{figure}

The \rjpsi single ratio is used to perform a number of additional cross-checks. The distribution of this ratio as a function of the angle between the leptons and the minimum \pt of the leptons is shown in Fig.~\ref{fig:rjpsi_differential1}, together with the spectra expected for the resonant and nonresonant decays.
No significant trend is observed in either \rjpsi distribution. Assuming the deviations observed are genuine mismodelling of the efficiencies, rather than statistical fluctuations, a total shift of \RK at a level less than $0.001$ would be expected due to these effects. This estimate takes into account the spectrum of the relevant variables in the nonresonant decay modes of interest and is compatible with the estimated systematic uncertainties on \RK. Similarly, the variations seen in \rjpsi as a function of all other reconstructed quantities examined are compatible with the systematic uncertainties assigned. In addition, \rjpsi is computed in two-dimensional intervals of reconstructed quantities, as shown in Fig.~\ref{fig:rjpsi_bin}. Again, no significant trend is seen.
 
\subsubsection*{Systematic uncertainties}

The majority of the sources of systematic uncertainty affect the relative efficiencies between nonresonant and resonant decays. These are included in the fit to \RK by allowing the relative efficiency to vary within Gaussian constraints. The width of the constraint is determined by adding the contributions from the different sources in quadrature. Correlations in the systematic uncertainties between different trigger categories and run periods are taken into account. Systematic uncertainties affecting the determination of the signal yield are assessed using pseudoexperiments generated with variations of the fit model. Pseudoexperiments are also used to assess the degree of bias originating from the fitting procedure. The bias is found to be 1\% of the statistical precision, \ie negligible with respect to other sources of systematic uncertainty.

For the nonresonant \BuKee decays, the systematic uncertainties are dominated by the modelling of the signal and background components used in the fit. The effect on \RK is at the 1\% level. A significant proportion (0.7\%) of this  uncertainty comes from the limited knowledge of the $K\pi$ spectrum in \BuBdKpiplusee decays. In addition, a 0.2\% systematic uncertainty is assigned for the potential contribution from partially reconstructed decays with two additional pions. 
A comparable uncertainty to that from the modelling of the signal and background components is induced by the limited sizes of calibration samples. Other sources of systematic uncertainty, such as the calibration of \Bu production kinematics, the trigger calibration and the determination of the particle identification efficiencies, contribute at the few-permille or permille level, depending strongly on the data-taking period and the trigger category.


The uncertainties on parameters used in the simulation model of the signal decays affect the \qsq distribution and hence the selection efficiency. These uncertainties are propagated to an uncertainty on \RK using predictions from the {\sc{flavio}} software package~\cite{Straub:2018kue} but give rise to a negligible effect. Similarly, the differing \qsq resolution between data and simulation, which alters estimates of the \qsq migration, has negligible impact on the result.

%% file: acknowledgements.tex
\section*{Acknowledgements}
%
%
\noindent We express our gratitude to our colleagues in the CERN
accelerator departments for the excellent performance of the LHC. We
thank the technical and administrative staff at the LHCb
institutes.
We acknowledge support from CERN and from the national agencies:
CAPES, CNPq, FAPERJ and FINEP (Brazil); 
MOST and NSFC (China); 
CNRS/IN2P3 (France); 
BMBF, DFG and MPG (Germany); 
INFN (Italy); 
NWO (Netherlands); 
MNiSW and NCN (Poland); 
MEN/IFA (Romania); 
MSHE (Russia); 
MICINN (Spain); 
SNSF and SER (Switzerland); 
NASU (Ukraine); 
STFC (United Kingdom); 
DOE NP and NSF (USA).
We acknowledge the computing resources that are provided by CERN, IN2P3
(France), KIT and DESY (Germany), INFN (Italy), SURF (Netherlands),
PIC (Spain), GridPP (United Kingdom), RRCKI and Yandex
LLC (Russia), CSCS (Switzerland), IFIN-HH (Romania), CBPF (Brazil),
PL-GRID (Poland) and NERSC (USA).
We are indebted to the communities behind the multiple open-source
software packages on which we depend.
Individual groups or members have received support from
ARC and ARDC (Australia);
AvH Foundation (Germany);
EPLANET, Marie Sk\l{}odowska-Curie Actions and ERC (European Union);
A*MIDEX, ANR, Labex P2IO and OCEVU, and R\'{e}gion Auvergne-Rh\^{o}ne-Alpes (France);
Key Research Program of Frontier Sciences of CAS, CAS PIFI, CAS CCEPP, 
Fundamental Research Funds for the Central Universities, 
and Sci. \& Tech. Program of Guangzhou (China);
RFBR, RSF and Yandex LLC (Russia);
GVA, XuntaGal and GENCAT (Spain);
the Leverhulme Trust, the Royal Society
 and UKRI (United Kingdom).

%% file: Authorship_LHCb-PAPER-2021-004.tex
\centerline
{\large\bf LHCb collaboration}
\begin
{flushleft}
\small
R.~Aaij$^{32}$,
C.~Abell{\'a}n~Beteta$^{50}$,
T.~Ackernley$^{60}$,
B.~Adeva$^{46}$,
M.~Adinolfi$^{54}$,
H.~Afsharnia$^{9}$,
C.A.~Aidala$^{85}$,
S.~Aiola$^{25}$,
Z.~Ajaltouni$^{9}$,
S.~Akar$^{65}$
% ...

{\footnotesize \it

% ...
$^{9}$Universit{\'e} Clermont Auvergne, CNRS/IN2P3, LPC, Clermont-Ferrand, France\\
% ...
$^{25}$INFN Sezione di Milano, Milano, Italy\\
% ...
$^{32}$Nikhef National Institute for Subatomic Physics, Amsterdam, Netherlands\\
% ...
$^{46}$Instituto Galego de F{\'\i}sica de Altas Enerx{\'\i}as (IGFAE), Universidade de Santiago de Compostela, Santiago de Compostela, Spain\\
% ...
$^{50}$Physik-Institut, Universit{\"a}t Z{\"u}rich, Z{\"u}rich, Switzerland\\
% ...
$^{54}$H.H. Wills Physics Laboratory, University of Bristol, Bristol, United Kingdom\\
% ...
$^{60}$Oliver Lodge Laboratory, University of Liverpool, Liverpool, United Kingdom\\
% ...
$^{65}$University of Cincinnati, Cincinnati, OH, United States\\
% ...
$^{68}$Syracuse University, Syracuse, NY, United States\\
% ...
$^{85}$University of Michigan, Ann Arbor, United States, associated to $^{68}$\\
\bigskip
% ...
\medskip
}
\end{flushleft}